\documentclass[notitlepage,a4paper,superscriptaddress,nofootinbib,showkeys,aps,pre]{revtex4-2}

\usepackage{amssymb,amsmath,amsfonts}
\usepackage{graphicx}

\usepackage{wrapfig, comment}
\usepackage{booktabs}
\usepackage{enumerate}
\usepackage{color}
\usepackage{url}
\usepackage{hyperref}
\usepackage{times}
\usepackage[normalem]{ulem}
\usepackage{xcolor}
\usepackage{tabularx}

\newcommand{\beq}{\begin{equation}}
\newcommand{\eeq}{\end{equation}}
\newcommand{\tTrain}{t_{\textrm{train}}}

\begin{document}

\title{Practical identifiability and parameter estimation of compartmental epidemiological models}

\author{Q. Y. Chen}
\email[Corresponding author; email address: ]{cqy@umass.edu}
\affiliation{Department of Mathematics and Statistics, University of Massachusetts,
Amherst, MA 01003-4515, USA}

\author{Z. Rapti}
\affiliation{Department of Mathematics and Carl R.Woese Institute for Genomic Biology, University of Illinois
at Urbana-Champaign}

\author{Yannis Drossinos}
\affiliation{Thermal Hydraulics \& Multiphase Flow Laboratory,
Institute of Nuclear \& Radiological Sciences and Technology, Energy \& Safety,
National Centre for Scientific Research ``Demokritos", 15314 Agia Paraskevi, Greece}

\author{J. Cuevas-Maraver}
\affiliation{Grupo de F\'{i}sica No Lineal, Departamento de F\'{i}sica Aplicada I,
Universidad de Sevilla. Escuela Polit\'{e}cnica Superior, C/ Virgen de \'{A}frica, 7, 41011-Sevilla, Spain \\
Instituto de Matem\'{a}ticas de la Universidad de Sevilla (IMUS). Edificio Celestino Mutis. Avda. Reina Mercedes s/n, 41012-Sevilla, Spain}

\author{G.A. Kevrekidis}
\affiliation{Department of Applied Mathematics and Statistics, Johns Hopkins University, Baltimore, MD 21218, USA}

\author{P. ~G. Kevrekidis}
\affiliation{Department of Mathematics and Statistics, University of Massachusetts,
Amherst, MA 01003-4515, USA}

\date{\today, \texttt{main\_v7.tex}}

\begin{abstract}
Practical parameter identifiability in ODE-based epidemiological models is a known issue, yet one that merits further study. It is essentially ubiquitous due to noise and errors in real data. In this study, to avoid uncertainty stemming from data of unknown quality, simulated data with added noise are used to investigate practical identifiability in two distinct epidemiological models. Particular emphasis is placed on the role of  initial conditions, which are assumed unknown, except those that are directly measured. Instead of just focusing on one method of estimation, we use and compare results from various broadly used methods, including maximum likelihood and Markov Chain Monte Carlo (MCMC) estimation.

Among other findings, our analysis revealed that the MCMC estimator is overall more robust than the point estimators considered.
Its estimates and predictions are improved when the initial conditions of certain compartments are fixed so that the model becomes
globally identifiable. For the point estimators,
whether fixing or fitting the that are not directly measured improves parameter estimates is model-dependent. 
Specifically, in the standard SEIR model, fixing the initial condition for the susceptible population $S(0)$ improved parameter estimates, while this was not true when fixing the initial condition of the asymptomatic population in a more involved model. 
Our study corroborates the change in quality of
parameter estimates upon usage of 
pre-peak or post-peak time-series under
consideration.
Finally, our examples suggest that in the presence of significantly noisy data, the value of structural identifiability is moot. 
\end{abstract}

\maketitle


\section{Introduction}
\label{sec:Intro}

Differential equation models containing many parameters are used extensively to model dynamical systems across various disciplines, extending, e.g., from ship-steering dynamics in ocean 
engineering all the
way to the spread of infectious diseases in epidemiology~\cite{Epidemic_Summary}. Some of these model parameters may not be directly measurable or a-priori known, hence they need to be estimated from existing observations. 

Although the parameter identifiability problem has been defined in multiple previous studies, we include it here for the sake of clarity. Consider a dynamical
system in the form
\begin{align}
\dot{\mathbf{x}}(t) = \mathbf{f}(\mathbf{x}(t); \Theta), \qquad \qquad
\mathbf{x}(0) = \mathbf{x}_0(\Theta) ,
\label{eq:dynsyst}
\end{align}
where vector $\mathbf{x}$ denotes the state variables,  $\Theta$ the model parameters and $\mathbf{x}_0$ the initial conditions of the state variables. We use $\mathbf{y}$ to denote the observations of the dynamical system
\begin{align}
    \mathbf{y}(t) = \mathbf{g}(\mathbf{x}(t),\Theta),
\label{eq:observ}
\end{align}
where the vector-valued function $\mathbf{g}$ describes the dependence of the observables on the state variables $\mathbf{x}$ and model parameters $\Theta$.  

 Parameters may be globally (uniquely) or locally identifiable, in which (latter) case a finite number of possible values for the parameter exist, all of which lead to the identical observables. 
A model such as (\ref{eq:dynsyst}) with observations (\ref{eq:observ}) is called structurally globally (locally) identifiable if all unknown parameters are globally (locally) identifiable \cite{Audoly2001}.
Structural identifiability analysis,  via differential algebra \cite{Audoly2001} or other methods \cite{Miao2011} is a theoretical analysis that assumes the observables (data)  are noise-free, continuous time series.

Practical identifiability,  on the other  hand, is linked  to the amount (e.g., length of the time-series for $\mathbf{y}$) and the quality of data. According to one definition of practical identifiability, a model parameter is practically identifiable, if a finite confidence interval for its estimate exists \cite{raue2009}. The motivation behind this is as follows. Given an objective function, such as the sum of squared errors between the model output and the data, a parameter is globally identifiable if its estimated value is a unique minimum of the objective function (termed loss function in following sections). However, for certain objective functions, although a minimum may exist, the confidence interval may be infinite, hence giving rise to practical non-identifiability. For a structurally globally identifiable  model (all model parameters are globally identifiable) that has some practically non-identifiable parameters, the issue can be resolved by reducing measurement errors and/or obtaining additional data points. In contrast, when the model is structurally non-identifiable (has one or more unidentifiable parameters) the issue can only be resolved by modifying the model or obtaining additional observed state variables.

Emerging epidemics (Ebola in 2014) \cite{tuncer2018} and pandemics (COVID-19) \cite{sauer2021} often revive the issue of identifiability and reliable estimation of model parameters. 
Recent works, motivated  by the COVID-19 pandemic, have often focused on structural identifiability \cite{chowell2023} and others investigate both structural and practical identifiability \cite{zhang2021}. While theoretical methods \cite{Audoly2001, Miao2011} and software \cite{daisy, combos, Pogudin_SIAN} exist for the study of structural parameter identifiability, practical parameter identifiability is not as well-studied. This is one of the  knowledge gaps that the present study is attempting  to address. We also investigate structural identifiability questions that remain open. For instance, as pointed out in~\cite{IC2003},  differential-algebra identifiability results may be incorrect for certain special initial conditions. Another pertinent point is the use of partial a-priori knowledge of parameter ranges informed by the underlying epidemiology. Usually, when an identifiability analysis is performed, no bounds are imposed on the parameters or initial conditions. We will show, through specific examples, that a locally identifiable model may become globally identifiable when certain constraints are enforced. 

In the next section, Section \ref{sec:Ident}, we present results on theoretical (structural) identifiability of parameters and initial conditions in two epidemiological models. In Section \ref{sec:Estimate}, we review commonly used  parameter estimation methods. In Section~\ref{sec:Predict},  we report numerical results and discuss the resulting practical identifiability issues. Section \ref{sec:Concl} offers a summary of our main findings and suggestions for future work. 

\section{Parameter Identifiability}
\label{sec:Ident}

In this section, we present our methodology for studying practical identifiability and sensitivity and determine the structural identifiability of parameters and initial conditions in two compartmental epidemiological ODE models. 

\subsection{Practical identifiability and sensitivity analysis}
\label{sec:Hessian}

There is no absolute definition of \emph{practical identifiability} \cite{raue2009}. The concept was introduced because often one is unable to obtain sharp estimates of some parameters even though they are globally identifiable. In essence,  theoretical identifiability results often do not necessarily carry over to practical calculations; for an example see \cite{Eisenberg2013}. 
One may argue that practical identifiability analysis is more meaningful
than structural global or local identifiability analysis, as all parameter
estimates are intrinsically approximations. Quite
often,  theoretical analyses ignore the consequences and implicit limitations of the complex algorithms used by the estimators (e.g. the optimization algorithm) that may not converge or finish 
in practical settings given a set of computer resources.

The quality of a parameter estimate is directly related to the sensitivity of the objective/loss function of the data (i.e., the observables) to that parameter. Note that in~\cite{Joubert2021} the authors proposed to use the singular values of the sensitivity of the output data, rather than any particular objective/loss function, to determine which parameters or initial conditions are not globally identifiable. Here,  instead, we calculate the sensitivity matrix when the loss (error) function is simply the sum of the squared error
$$
L(\mathbf{\Theta}) = \sum_i [D(t_i, \mathbf{ \Theta}) - D(t_i, \mathbf{ \Theta^*})]^2 ,
$$
where $D(t_j, \mathbf{ \Theta})$ is the model output for $p$ given parameters and initial conditions 
$\mathbf{\Theta} = (\theta_1,\ldots, \theta_p)$
at time $t_j$,
$D(t_j, \mathbf{ \Theta^*})$ is the observed time-series data, and  $\mathbf{ \Theta^*}$ is the exact parameter and initial condition
set  that lead to these data. The goal is to estimate  $\mathbf{ \Theta^*}$ by minimizing the loss function. 
Note that this loss function is not identical to the loss function
of the estimators discussed in Sec.~\ref{sec:Estimate} and used in Sec.~\ref{sec:Predict}.
Regardless, its use can inform on the sensitivity of the model output to model parameters and initial conditions.

The Taylor expansion of the loss function near  $\mathbf{ \Theta^*}$ takes the form:
\begin{equation*}
L(\mathbf{\Theta}) = \frac{1}{2!} (\mathbf{\Theta}-\mathbf{\Theta}^*)^T  \cdot
\left . \mathbf{H} \right |_{\mathbf{\Theta}^*} \cdot (\mathbf{\Theta}-\mathbf{\Theta}^*) + \text{high-order terms},
\end{equation*}
since $L(\mathbf{\Theta}^*) = 0$, and the first derivative vanishes at the extremum.
The elements $h_{jk}$ of the Hessian $\mathbf{H}$ evaluated at $\mathbf{\Theta}^*$ are
$$
h_{jk} = \left . \frac{\partial^2 L}{\partial \theta_j \partial \theta_k}\right |_{\mathbf{\Theta}^*}  = \left. 2 \sum_i \frac{\partial  D(t_i, \mathbf{\Theta})}{\partial \theta_j} \cdot \frac{ \partial D(t_i, \mathbf{\Theta})}{\partial \theta_k}
\right|_{\mathbf{\Theta}^*}
$$
The eigenvalues of the Hessian measure the sensitivity of the loss function to changes of the parameters (and initial conditions) along each eigendirection. 
A large difference between eigenvalues indicates that the data $D$, or more precisely the loss function, depend weakly on  certain parameters or initial conditions. In particular,  a small, tending to zero eigenvalue indicates that changes of the associated parameters will not lead to significant changes of the loss function (and the predicted data $D(t)$). 
However, the eigenvalues do not provide information on local identifiability.
If the gap between eigenvalues is very large, the system is
called a sloppy system~\cite{Sloppy2007}.

To calculate the Hessian matrix, we need to compute the derivative of state variables with respect to parameters and initial conditions.
For the generic system (\ref{eq:dynsyst}), the derivative of state variables $\mathbf{x}$ with respect to parameters $\Theta$ and initial conditions $\mathbf{x}_0$ satisfy the equations
$$
\frac{d}{dt}\left[\frac{\partial \mathbf{x}}{\partial \Theta} \right] =  \nabla \mathbf{f} \; \frac{\partial \mathbf{x}}{\partial \Theta}
      + \frac{\partial \mathbf{f}}{\partial \Theta},
\qquad \qquad 
\frac{d}{dt}\left[\frac{\partial \mathbf{x}}{\partial \mathbf{x}_0} \right] =  \nabla \mathbf{f} \; \frac{\partial \mathbf{x}}{\partial \mathbf{x}_0}.
$$
The initial conditions for $\partial \mathbf{x}/\partial \Theta$ 
may be approximated by a first order finite difference, while the initial conditions for 
$\partial \mathbf{x}/ \partial \mathbf{x}_0$
is simply the identity matrix. 
Reference~\cite{Mexico} presents a calculation of the Hessian of the norm in a compartmental epidemiological model as previously described.

\subsection{Identifiability analysis of compartmental epidemiological models}
\label{sec:Input}

For the identifiability analysis, we initially rewrite the model equations by expressing as many as possible state variables as functions of the observable $D(t)$. We, then, use the Julia package StructuralIdentifiability~\cite{Pogudin_2022} to find the identifiable parameters, and finally we use the SIAN~\cite{Pogudin_SIAN}  WebApp  software to find identifiable combinations of the parameters and to assess the identifiability of the initial conditions.

\subsubsection{Susceptible-Exposed-Infected-Recovered (SEIR) model}

The Susceptible-Exposed-Infected-Recovered (SEIR) compartmental
epidemiological model
\begin{subequations}
\begin{eqnarray}
S' &=&  -\beta S I ,  \label{eq:SEIR_S} \\
E' &=& \beta SI - \sigma E  ,  \label{eq:SEIR_E}  \\
I'  &=& \sigma E - \gamma I ,  \label{eq:SEIR_I} \\
R' &=& \gamma I ,
\end{eqnarray} 
\end{subequations}
has been widely studied. 
Assume the observable (i.e., the data) is $I(t)$. Following the approach in~\cite{Eisenberg2013}, we obtain
$$
I I''' - I' I'' + (\gamma + \sigma) I I'' - (\gamma + \sigma) (I')^2 + \beta I^2 I'' + \beta(\gamma + \sigma) I^2 I' + 
\beta \gamma \sigma I^3 =0 .
$$
Consequently, the identifiable parameters are
\beq
\beta, \quad \gamma \sigma, \quad \gamma + \sigma,
\label{eq:Symmetry}
\eeq
namely, $\beta$ is globally, and $\gamma, \sigma$ are locally identifiable.
A more complete result is obtained using the SIAN~\cite{Pogudin_SIAN} WebApp software:
$$
\beta: \; \text{globally identifiable}; \qquad\quad  \gamma, \sigma, S(0), E(0): \; \text{locally identifiable}.
$$
Moreover, all the locally identifiable parameters and initial conditions have two possible solutions, either of which may be obtained
by estimators (cf. Sec.~\ref{sec:Estimate}).
Consider the following two sets of parameters and initial conditions, identified by the superscript $(1,2)$,
that lead to the same observable $I(t)$:
$$
\{\beta^{(1)}, \sigma^{(1)}, \gamma^{(1)}, S^{(1)}(0), E^{(1)}(0), I^{(1)}(0), R^{(1)}(0)\}, \qquad  \text{and} 
\qquad \{ \beta^{(2)}, \sigma^{(2)}, \gamma^{(2)}, S^{(2)}(0), E^{(2)}(0), I^{(2)}(0), R^{(2)}(0) \}.
$$
Let $N$ be the conserved total population: $N = S(0) + E(0) + I(0) + R(0)$.
These two sets share  the same $\beta$,  as it is globally identifiable,
and $I(0)$, as it is a datum (the observable at $t=0$).
The second set has $\gamma$ and $\sigma$ swapped due to the dual symmetry
identified in the identifiability analysis, see Eq.~(\ref{eq:Symmetry}).
Hence,
\beq
\beta^{(2)} = \beta^{(1)}, I^{(2)}(0)= I^{(1)}(0), \sigma^{(2)} = \gamma^{(1)}, 
\gamma^{(2)} = \sigma^{(1)}, R^{(2)}(0) = N - S^{(2)}(0) - E^{(2)}(0) - I^{(2)}(0) .
\label{eq:SymmetrySEIR}
\eeq
Lastly, the initial conditions $S^{(2)}(0)$ and $E^{(2)}(0)$ may be derived from the original system as follows.
We determine $E^{(2)}(0)$ from the $I'$ equation, Eq.~(\ref{eq:SEIR_I}), under the two sets of parameters
and taking the $t \to 0$ limit:
$$
\begin{cases} I' & = \sigma^{(1)} E^{(1)} -\gamma^{(1)} I \\
I' &= \sigma^{(2)} E^{(2)} - \gamma ^{(2)} I \end{cases} \quad
\xrightarrow[t \to 0]{} 
\begin{cases} I'(0) &= \sigma^{(1)} E^{(1)}(0) - \gamma^{(1)} I(0) \\E^{(2)}(0) &= \frac{1}{\sigma^{(2)}} (I'(0) + \gamma^{(2)} I(0))
= \frac{\sigma^{(1)}}{\sigma^{(2)} }E^{(1)}(0) + \frac{\gamma^{(2)}  - \gamma^{(1)}}{\sigma^{(2)} } I(0) . \end{cases}
$$
For $S^{(2)} (0)$, we start with the $E'$ equation, Eq.~(\ref{eq:SEIR_E}) to obtain
$$
S = \frac{1}{\beta I}(E' + \sigma E) = \frac{1}{\beta \sigma I} \left [I'' + (\gamma + \sigma)I' + \gamma \sigma I\right] .
$$
Consequently,
$$
S^{(1)} (\beta^{(1)} \sigma^{(1)} ) = S^{(2)}  (\beta^{(2)}  \sigma^{(2)} ) 
\quad \xrightarrow[t \to 0]{} \; S^{(2)}(0) = S^{(1)}(0) \frac{\sigma^{(1)}}{\sigma^{(2)}}.
$$
Since $\gamma, \sigma, S(0)$ and $E(0)$ are locally identifiable, 
either of the two possible values may be obtained by the estimators. 
The common practice to enforce bounds for every parameter and initial condition may effectively make everything globally identifiable, as the bounds may eliminate the scenario of multiple possible values. The specification of any of these locally identifiable parameters or initial conditions to its exact value can also make everything globally identifiable. 

\subsubsection{ Exposed-Asymptomatic-Infected-Hospitalized-Recovered-Dead (EAIHRD) model}
\label{sec:IdentifyFokas}

As previously discussed, the differential-algebra approach  to identifiability analysis~\cite{Eisenberg2013} is based on expressing the dynamics of the whole system as a single high-order ODE of the observable. When the model is complex, such a task may be unattainable, even with the help of symbolic computing software, like Mathematica. Here,  we also consider an alternative approach to study the identifiability of a rather complex epidemiological compartmental model with the Julia package StructuralIdentifiability~\cite{Pogudin_2022}.

We analyze the one-age Exposed-Asymptomatic-Infected-Hospitalized-Recovered-Dead (EAIHRD) model presented in~\cite{Fokas2020}, and described by the following ODEs
\begin{subequations} \label{eq:CSF}
\begin{eqnarray}
A' &= & a E - r_1 A ,  \label{eq:CSF_A} \\
I' &= & sE - (h+r_2) I  ,  \label{eq:CSF_I} \\
H' &= & h I - (r_3+d) H  ,  \label{eq:CSF_H} \\
R' &= & r_1 A + r_2 I + r_3 H  ,  \label{eq:CSF_R} \\
D' &= & d H  ,  \label{eq:CSF_D} \\
E' &= & \left [ N - (A+I+H+R+D+E) \right] (c_1 A + c_2 I) - (a+s) E  , \label{eq:CSF_E} 
\end{eqnarray}
\end{subequations}
where as before $I$ is the infected compartment, $E$ is the exposed, and $R$ is the recovered compartment.The additional populations are the asymptomatic population $A$ (infected, infectious, without symptoms), $H$ the hospitalized  population, and $D$ the permanently removed population due to fatality. (Note that in~\cite{Fokas2020}, $S$ is the infected compartment (i.e., sick) and $I$ plays the role of
the exposed in the above discussion.) $N$ is the total population, which is assumed to be a constant. The model has nine parameters: $a, s, r_1, r_2, r_3, h, d, c_1,c_2$. The state variables are not normalized by the total population, therefore $c_1, c_2$
(the transmission rates) will be very small. The observable is  $D(t)$,  the cumulative number of recorded fatalities at day $t$.

Even though the model is moderately complex the Julia package StructuralIdentifiability was able to complete the relevant task. The only globally identifiable parameter is $r_1$, the rest are all non-identifiable (not even locally identifiable). 
On the other hand, the SIAN WebApp is unable to  produce any results. Consequently, we do not have any information on which initial conditions are identifiable and what combinations of the parameters and initial conditions are locally or globally identifiable.

To gain more insight on the identifiability of the model following the differential algebra approach to identifiability analysis one may rewrite the system as a single high-order ODE of the observable. The new high-order equation will naturally have new parameters as coefficients that 
would be combinations of the original model parameters. 
One such result was obtained in~\cite{Fokas2020}. Then with the aid of the package StructuralIdentifiability, we indeed obtained the full identifiability results. However, the new high-order ODE contains a logarithmic term that eventually makes it impossible for any of the estimators discussed in Sec.~\ref{sec:Estimate} to finish, unless the initial parameter guess in the estimation (or optimization) procedure is very close to its exact value.
 
We thus follow a slightly different approach which gives us the same identifiability results. This approach is more general and can be applied  to various models, such as the two-age model in~\cite{Fokas_2Age} and the SARS-CoV-2 omicron-variant model with vaccination~\cite{Vaccine}. The main idea is to stop short of writing the system as a single differential equation. For moderately-complex systems, the single differential equation will be extremely complex, and may have hundreds of terms due to the nonlinear terms of the model. Our strategy is to rewrite the original system as a high-order differential equation for the observable, coupled with one additional differential equation for a state variable involved in the nonlinear term ($A$ in this case). Then,  StructuralIdentifiability may be used to perform the identifiability analysis.

We introduce new combinations of parameters, identical to those in~\cite{Fokas2020}:
\begin{align*}
& F = a+s, \quad R_2 = r_2 +h, \quad R_3 = r_3 + d,\quad  C_1 = \frac{ac_1}{hsd}, \quad C_2 = \frac{sc_2}{hsd}, \\
& k_1 = F R_2 R_3, \quad  k_2 = FR_2 + F R_3 + R_2R_3, \quad k_3 = F + R_2 + R_3.
\end{align*}
Then, we use the model equations to rewrite state variables $H, I$ and $E$ as functions of $D$ and its derivatives:
\begin{subequations} \label{eq:HIE}
\begin{eqnarray}
H &= & \frac{D'}{d} , \\
I &= & \frac{1}{hd} (D'' + R_3 D') ,  \\ 
E&= & \frac{1}{hsd} \, \left [D''' + D'' (R_2+R_3) + D' R_2 R_3  \right ]  ,
\end{eqnarray}
\end{subequations}
to obtain
\begin{eqnarray*}
E + F \int_0^t  E(\tau) d \tau &=&  \frac{D''' + k_3 D'' + k_2 D' + k_1 D}{hsd} - \frac{F}{hsd} \left[ D'' + D' (R_2+R_3)+D R_2 R_3 \right]_{t=0}.
\end{eqnarray*}
Subsequently,  we add the first 5 equations Eqs.~(\ref{eq:CSF_A} - \ref{eq:CSF_D}):
\begin{eqnarray*}
\frac{d}{dt} ( A + I + H + R + D) = F \; E  \quad \Rightarrow
A+I+H+R+D = F \int_0^t E(\tau) d \tau +  \left[A+I+H+R+D \right]_{t=0},
\label{eq:Add}
\end{eqnarray*}
substitute the resulting equation into Eq.~(\ref{eq:CSF_E}),
multiply it by $hsd$, to obtain 
\begin{equation*}
\frac{ d Q_3}{ dt}  = -Q_3 (c_1 A  + c_2 I) , ~~~~\mbox{where} ~~~
Q_3 = D''' + k_3 D'' + k_2 D' + k_1 D - \alpha ,
\end{equation*}
with
\begin{equation}
\alpha \equiv  hsd \left[N -(A+I+H+R+D) \right]_{t=0}  +  F \left[ D'' + D' (R_2+R_3)+D R_2 R_3 \right]_{t=0}  .
\label{eq:4thA_alpha}
\end{equation}

The new system now takes the form
\begin{eqnarray*}
\frac{d A}{d t} &=& a E - r_1 A ,  \\
\frac{ d Q_3}{ dt} & =& -Q_3 (c_1 A  + c_2 I) .
\end{eqnarray*}
We use only the variables $A$ and $D$, and the derivatives of $D$
to rewrite the system as:
\begin{eqnarray*}
\frac{d A}{d t} &=& \frac{a}{hsd} \big [ D''' +D'' (R_2 + R_3) + D' R_2 R_3 \big ] - r_1 A , \\
D^{(4)} &=& - \left ( k_3 D''' + k_2 D'' + k_1 D' \right) -  \left( D'''+k_3 D'' + k_2 D' + k_1 D - \alpha \right) 
\left[ c_1 A + \frac{c_2}{hd} (D'' + R_3 D') \right] .
\end{eqnarray*}
Finally, by scaling $ A \frac{hsd}{a} = \tilde{A}$ we obtain
\begin{subequations} \label{eq:CSF4thA}
\begin{eqnarray}
\frac{d \tilde{A}}{d t} &= & [D''' +D'' (R_2 + R_3) + D' R_2 R_3] - r_1 \tilde{A} , \\
D^{(4)} &=& - \left( k_3 D''' + k_2 D'' + k_1 D'\right) -  \left(D'''+ k_3 D'' + k_2 D' + k_1 D - \alpha \right) \left [  C_1 \tilde{A} + C_2 (D'' + R_3 D') \right ] .
\end{eqnarray}
\end{subequations}
The new form of the model Eqs.~(\ref{eq:CSF}) has now seven parameters, $\alpha$ was defined  in Eq.~\eqref{eq:4thA_alpha}:
\begin{equation}\label{eq:4thA_para}
F = a+s, \quad R_2 = r_2 +h, \quad R_3 = r_3 + d,\quad  C_1 = \frac{ac_1}{hsd}, \quad C_2 = \frac{sc_2}{hsd},  \quad r_1, 
\quad \alpha.
\end{equation}
The model form described by Eqs.~\eqref{eq:CSF4thA} may be analyzed by the identifiability packages after rewriting
its as a first-order system. It turns out that
\begin{equation*}
R_3, \quad \alpha, \quad r_1,\quad  F + R_2,\quad F R_2, \quad  \frac{F-r_1}{C_1}, \quad  R_2 - C_2 \frac{F-r_1}{C_1}, 
\end{equation*}
are globally identifiable, and 
$$
F, \quad R_2, \quad C_1, \quad C_2, \quad \tilde{A}(0)
$$ 
are locally identifiable. Each of the locally identifiable parameters also has two possible solutions.
Hence, as in the case of the SEIR model, we find two sets of parameters and initial conditions that lead to identical $D(t)$.

The two sets of parameters and initial conditions that lead to the same $D(t)$ are shown in Table~\ref{tab:CSF_SetPar}. 
Note that we dropped the superscripts for the first set, and wrote the second set as a function of the first set. 

\begin{table}[hbtp]
\caption{Sets of parameters and initial conditions that
lead to the same observable $D(t)$.}
\label{tab:CSF_SetPar}
\begin{tabular}{ccccccccc} \\ \hline \hline
1st set & $F$ & $R_2$ & $R_3$ & $C_1,$  & $C_2$ & $r_1$ & $\alpha$  & $\tilde{A}(0)$ \\
2nd set & $R_2$ & $F$ & $R_3$ & $C_1 \frac{R_2-r_1}{F-r_1}$ & $C_2 + C_1 \frac{F-R_2}{F-r_1}$ & $r_1$ & $\alpha$ & 
$\frac{F-r_1}{R_2-r1}\tilde{A}(0) + \frac{F-R_2}{r_1-R_2} \left [ D''(0) + R_3 D'(0) \right ]$ \\ \hline \hline
\end{tabular}
\end{table}
If we set any locally identifiable parameter or initial condition
$[F, R_2, C_1, C_2$ or $\tilde{A}(0) ]$ to its exact value, or we enforce
bounds on some of them, the system may effectively become globally
identifiable.

\section{Parameter estimation methods}
\label{sec:Estimate}

We start the relevant discussion by briefly reviewing the methods we used to estimate model parameters and initial conditions. We denote $\{d_i\}$ the reported/collected data of certain variable(s) -- usually a time series of state variable(s). The time  series  $\{\tilde{d}_i\}$ denotes the state variables calculated from the epidemiological model. The model usually involves a certain number of parameters, the values of which are to be estimated. We also denote $\{\hat{d}_i\}$ as the exact time series of the state variable(s), which is often unknown.  As before, the vector $\mathbf{\Theta}$ consists of all parameters and initial conditions
that are to be estimated. In the following, the term parameters may refer to the parameters only or  both the parameters and the initial conditions that are to be estimated depending on the context.

We primarily used four estimators: Deterministic Optimization (DO), Maximum Likelihood Estimation (MLE),
Maximum \emph{A Posteriori} estimation (MAP), and a Bayesian Markov Chain Monte Carlo (MCMC). These
are described in the following subsections.  We also used the Ensemble Kalman Inversion estimator (EnKI)\cite{KodyEnKI} in 
two test cases, but we found its predictions to be less relevant for the purposes of our
comparison and decided not to include the corresponding results.

\subsection{Deterministic Optimization (DO)}
The deterministic optimization (DO) approach assumes that each parameter is a deterministic quantity. In particular,
it finds the values of the parameters by minimizing
certain deterministic ``error" between the actual data
and the numerical solutions of the model. (In the literature, this method is often called the Least Square Optimization.)
One particular error function (i.e., cost/objective function) is~\cite{barbastathis}:
\begin{equation} \label{eq:logobject}
L(\mathbf{\Theta}) = \sum_{i=1}^n | \log d_i - \log \tilde{d}_i|^2
\end{equation}
When $d_i$ and $\tilde{d}_i$ are close, the Taylor expansion
$\log d_i - \log \tilde{d}_i = d_i/\tilde{d}_i -1$
leads to another error function:
\begin{equation} \label{eq:object}
L(\mathbf{\Theta}) = \sum_{i=1}^n \left | \frac{d_i}{\tilde{d}_i} -1 \right|^2,
\end{equation}
which is more intuitive than Eq.~\eqref{eq:logobject}
because we simply minimize the sum of the squares of relative errors. 
Note that $L^1$ norm may also be used, but in this study we focus on the $L^2$ norm because it can be linked to the Maximum Likelihood Estimation (MLE) approach (discussed in Sec.~\ref{sec:MLE}).

Complex models often involve a large number of parameters,
rendering challenging the identification of the optimizer for any
optimization method. Even worse, there is no guarantee for the uniqueness of  such optimizer. 
To alleviate the multiple local minima problem,
and to obtain one optimizer (i.e., one set of parameter values and initial conditions), 
we choose an \emph{ad hoc} and commonly used approach: 
we run the optimization procedure a few times (10 in our study) with random initial guesses,
and we take the optimizer that leads to the least error as our estimate for the parameters. 

The DO is a point estimator: it finds a particular estimate for each parameter.
To make viable predictions with confidence intervals (or percentiles), 
the uncertainty/errors present in the data collecting procedure has to be incorporated.
One way is the bootstrapping procedure, the details of which are given below.
Another way is to assume that each parameter itself features a distribution (e.g., Bayesian estimator) 
instead of taking one particular value. Then we can obtain the distribution of each parameter, from which one can calculate the confidence intervals, 
prediction intervals etc.
Herein, we employ the bootstrapping procedure as presented in~\cite{chowell2019} for all the point estimators (DO, MLE, and MAP)
we use. 
It consists of the following steps:
\begin{itemize}
\item Find the ``optimal" parameter set by the optimization procedure. (The set of parameters with the least error 
among 10 different initial guesses will be chosen as the optimal values.)
\item Plug the ``optimal" parameter set into the model, and treat the model output as the ``numerical-truth", i.e., 
$\{\hat{d}_i\}$,
\item  Follow  an assumed noise structure (e.g.,~\eqref{eq:normal}) to generate $k$ realizations
$D_j = \{d_{i}^{(j)} \}_{i=1}^n, \; j=1,\ldots, k$. The superscript $(j)$ indicates the $j$-th generated time series, and 
will be omitted whenever there is no ambiguity.
\item Each realization $D_j = \{d_{i}^{(j)}\}_{i=1}^n$ is treated as one ``reported data set", and the same optimization procedure will be applied
to find one optimal set of parameter values.
The credible intervals and other statistics for each parameter can be then obtained from these $k$ sets. 
\end{itemize}

The bootstrapping procedure assumes a noise structure and noise level for the reported data,
which are completely independent from the optimization procedure, and can be chosen empirically.
We compared the normal, Poisson,  negative binomial, and log-normal noise structures~\cite{chowell2017} and did not find fundamental 
differences among the results. Subsequently, we only include the results for the normal noise structure.

The above bootstrapping is a regularization in a certain way. For optimization problems, the data are often polluted with noise.
Bootstrapping tackles this problem by simply assuming that the data
are not the truth but a noisy realization of the truth.
By including many realizations of the truth, it avoids over-fitting the original data. 

Reported data are assumed to be distributed according to
\begin{equation}\label{eq:normal}
\frac{d_i}{ \hat{d}_i} = 1 + \varepsilon_i, \quad \text{where i.i.d.} \; \varepsilon_i \sim N(0, \sigma_{\varepsilon}^2) ,
\end{equation}
namely, the noise is normally distributed.
We consider that on average, each unit of $\hat{d}_i$ will be counted as $(1+\varepsilon_i)$, where $\varepsilon_i$
satisfies the normal distribution $N(0, \sigma_{\varepsilon}^2)$. 
To reduce the number of parameters,  we assumed a single variance $\sigma_{\varepsilon}^2$, 
i.e,  the variance is independent of the time index $i$.
The noise level,  $\sigma_{\varepsilon}$, is  a newly introduced parameter associated with the noise structure, and can be chosen empirically. 
For certain data, it may be more appropriate to impose the noise structure on the daily incidence (i.e., the number of new daily cases), 
\begin{equation}\label{eq:DeltaNoise}
\frac{\Delta d_i}{ \Delta \hat{d}_i} = 1 + \varepsilon_i, \quad \text{where i.i.d.} \; \varepsilon_i \sim N(0, \sigma_{\varepsilon}^2), \qquad
\Delta d_i  = d_i-d_{i-1}, \;\; \Delta \hat{d}_i = \hat{d}_i - \hat{d}_{i-1}
\end{equation}
 
\subsection{Maximum Likelihood Estimation (MLE)} 
\label{sec:MLE}

Alternatively, one can find the ``optimal'' parameter values by maximizing
the likelihood function. The basic idea is to assume that the reported data satisfy certain distribution, and 
the optimal parameters are the ones that maximize the likelihood of the reported/observed data.
 The reported data are the one with the largest likelihood - most likely to happen.
Even though it is from a quite different perspective, the parameters are still deterministic and unique if globally identifiable,
i.e., MLE is still a point estimator.
The search for the optimal parameter set is again an optimization problem: the maximization of the likelihood function.
Hence, the same optimization approach as in the deterministic optimization may be used. 
The only change is the objective function: now we need to maximize the likelihood (log-likelihood actually)
instead of minimizing the fitting error.

{\bf Normal likelihood:} For a given set of parameter values, the model results are $\{\tilde{d}_i\}$. 
We assume the reported data $\{d_i\}$ satisfy the distribution given in Eq.~\eqref{eq:normal} (with $\hat{d}_i$ replaced by
$\tilde{d}_i$). 
Then, the likelihood function for observing $D = (d_1,\ldots, d_n)$ will be
\begin{equation}\label{eq:likelihood}
l(\mathbf{\Theta}) = \prod_{i=1}^n \frac{1}{\sqrt{2 \pi} ( \tilde{d}_i \sigma_L) } e^{- \frac{( d_i - \tilde{d}_i)^2}{2 \left(
\tilde{d}_i \sigma_L\right)^2}}
\end{equation}
because
\begin{equation}\label{eq:deltaDnormal}
d_i \sim  \tilde{d}_i \, N(1, \sigma_L^2) = N( \tilde{d}_i, ( \tilde{d}_i \,\sigma_L)^2).
\end{equation}
Maximizing the likelihood is equivalent to finding the minimum of 
the negative logarithm of the likelihood function. That is, the objective function is now:
\begin{equation} \label{eq:MleNormal}
L(\mathbf{\Theta}) = - \ln l(\mathbf{\Theta})  \quad \sim \quad  n \ln \sigma_L + \sum_{i=1}^n \ln  \tilde{d}_i + \frac{1}{2\sigma_L^2} 
\sum_{i=1}^n \left( \frac{ d_i}{ \tilde{d}_i} -1 \right)^2
\end{equation}
where the constant terms were dropped from the final formula, as they do not affect the optimizer.
The last term is like the objective function used in the deterministic optimization approach.
 The extra parameter $\sigma_L$ will be treated as an additional parameter which will be estimated from the optimization procedure.
Similar to the noise structure in the bootstrapping procedure, the terms $d_i$ and $\tilde{d}_i$ in the likelihood function
 may be replaced by their corresponding daily new cases
$\Delta d_i$ and $\Delta \tilde{d}_i$.

Since MLE is a point estimator, we will employ the bootstrapping procedure to obtain percentiles for the parameter estimates and 
predictions of the state variables.
Note that the noise level $\sigma_{\varepsilon}$ in the bootstrapping can not be obtained from the above optimization procedure. It should be selected 
empirically or by other means.

\subsection{Maximum \emph{a posteriori} estimation (MAP)}

 According to Bayes' rule, the \emph{a posteriori} distribution of $\mathbf{\Theta}$ is:
\begin{equation}\label{eq:post}
\pi(\mathbf{\Theta}|D) = \frac{\pi(D|\mathbf{\Theta}) \cdot \pi(\mathbf{\Theta})}{\pi(D)}
\end{equation}
where $\pi(\mathbf{\Theta})$ is the prior distribution of the parameters and
$\pi(D|\mathbf{\Theta})$ is the likelihood function (e.g., $l(\mathbf{\Theta})$ in Eq.~\eqref{eq:likelihood}).
Instead of maximizing the likelihood, MAP seeks the parameter values that maximize the posterior distribution.
Similar to MLE, the objective function for MAP takes the form (assuming normal likelihood)
\begin{equation} \label{eq:MAPNormal}
L(\mathbf{\Theta}) = - \ln \pi(\mathbf{\Theta}|D)  \quad \sim \quad  n \ln \sigma_L + \sum_{i=1}^n \ln  \tilde{d}_i + \frac{1}{2\sigma_L^2} 
\sum_{i=1}^n \left( \frac{ d_i}{ \tilde{d}_i} -1 \right)^2 - \ln \pi(\mathbf{\Theta}) .
\end{equation}
Again, the constant terms were dropped since they do not affect the optimization at all. 
The advantage of MAP is the inclusion of the prior distribution (i.e., prior knowledge of the parameters), which serves
as regularization for the optimization procedure. 

For a fair comparison between estimators, we do not assume any advanced a priori knowledge for the parameters. 
The prior distribution of the $j$-th parameter $\Theta_j$ is simply assumed to satisfy
\begin{equation*}
\pi(\Theta_j) \sim N \left[ \frac{lb_j + ub_j}{2},   \left(\frac{ub_j-lb_j}{6}\right)^2 \right], 
\end{equation*}
where $[lb_j, ub_j]$ represents the bounds enforced for $\Theta_j$ in all estimators. Basically, we assume that its prior is a normal
distribution with the interval center as its mean, and in the interval it has 3 times standard deviation on either side of its mean.
If additional knowledge is available about the prior distribution, it can be  included as well.

Like MLE and DO estimators, MAP is still a point estimator, even though it starts off of the Bayesian framework.
Consequently, we will use the bootstrapping procedure to obtain the percentiles for
the parameter estimates and prediction of the state variables.

\subsection{Bayesian Estimation (Markov Chain Monte Carlo, MCMC)}
In the Bayesian approach, the parameters or initial conditions do not take a deterministic value. 
Instead, every parameter or initial condition to be estimated features a distribution.
The goal (and outcome) is to generate a list (chain) of values for each parameter or initial condition according to 
the posterior distribution $\pi(\mathbf{\Theta}|D)$ given in Eq.~\eqref{eq:post}. 
 Usually the logarithm of the posterior density
function is needed to generate such a chain.
When assuming that the errors for each data point $d_i$  are independent, 
the normal likelihood function $\pi(D|\mathbf{\Theta}) $ is given in Eq.~\eqref{eq:likelihood}, and 
the negative logarithm of the posterior density function is given in Eq.~\eqref{eq:MAPNormal}.
Note that the dropped constant terms play no role in the sampling process.

We employ a Metropolis-Hastings type algorithm~\cite{Metropolis}, named $t$-walk~\cite{twalk} to sample from the above
posterior distribution. 
In particular, the $t$-walk maintains two independent points in the sample space, consequently two chains will be obtained.
The proposal distribution for the new points (i.e. 'moves') are chosen such that the algorithm is invariant to scale,
and approximately invariant to affine transformations of the state space. Then a standard Metropolis-Hastings acceptance/rejection
procedure is followed.
The potential scale reduction factor (PSRF)~\cite{PSRF_GB, PSRF_Vats} is calculated 
to ensure the MCMC chain achieves the required accuracy for a prescribed confidence level. 
Upon the convergence, we will have a MCMC chain, each point of which is a sample of all the parameters.
The chain is usually very long, especially when issues of non-identifiability are present.
In order to calculate the confidence intervals or percentiles of each parameter and the state variable(s) prediction, 
we only utilize a certain number of values 
from the chain in a conventional way: start backward from the end of the chain, and select one point among every few (or dozens of) points.
This process is called skipping, and is necessary as the consecutive points on the chain will be correlated. However, the common ``burn-in'' period is not needed since in
our simulations there is always a very large percentage portion of the chain that is not utilized, simply because the chain is too long. Unfortunately, there is no rule to specify how long the skipping step should be.

Confidence intervals or percentiles of the parameters and prediction can be readily computed as a chain of parameter values is obtained. 
No bootstrapping is required.

\section{Parameter estimation and prediction of state variables}
\label{sec:Predict}

We apply the estimators summarized in the previous section to the two models of Sec.~\ref{sec:Input}. The goal is to evaluate the quality of the estimated parameters and the predicted time series of the state variable(s) for different estimators under various conditions. We decided not to use actual data. Instead, we initially assume a typical value for each parameter and initial condition. With these values the model generates a synthetic time series of the observable(s).  This synthetic time series (with noise added to make it more realistic) is then treated as the ``reported" (i.e., actual) data that are used to estimate the parameters and initial conditions. The noise added to the original synthetic series $\{d_i\}$ was chosen to be
$$
\{d_i\} \quad \Rightarrow \quad \{ d_i \cdot N(1, \hat{\sigma}^2_{\varepsilon})\}, 
$$
i.e., a normal noise of relative size $\hat{\sigma}_{\varepsilon}$.  It may be more appropriate to add the noise to the daily new incidence, depending on the model and the data. Note that the noise level $\hat{\sigma}_{\varepsilon}$ is added to make the tests more realistic: there is no need for such artificial noise if we deal with actually reported data. At the same time, the noise level $\sigma_{\varepsilon}$ in the bootstrapping procedure is obtained empirically in an effort to capture the noise in the reported data.
For all the test cases, 200 sets of parameter/initial condition estimates were obtained, unless otherwise specified.

The algorithm used to estimate parameters and initial conditions involves choosing a training period, defined by the time $\tTrain$. The estimators use data for that period only, i.e.,  data within the period [$t_{\textrm{init}}, t_{\textrm{init}} + \tTrain$] with $ t_{\textrm{init}}$ the initial time of the data, to estimate the parameters. Subsequently, we  used the optimized parameters to calculate the predicted $5\%-95\%$ variation of the state variables, or the  box plot percentiles, as described in Section~\ref{sec:Estimate}. In addition to presenting model predictions for the observable, we also chose to compare predictions for a non-observable state variable. We wanted to investigate whether in a system possibly globally identifiable, but practically without sharp parameter estimates, a reasonable prediction of the observable also implied a reasonable prediction of a non-observable. Equivalently,  the question addressed is if the parameter estimates are not sharp and the observable prediction good, does that imply 
a potential expectation 
that model predictions for a non-observable would be equally accurate?

\subsection{Susceptible-Exposed-Infected-Recovered (SEIR) model}
\label{sec:predict_SEIR}

According to the identifiability analysis,  Sec.~\ref{sec:Input}, parameters $\sigma, \gamma, S(0), E(0)$ are locally identifiable. To estimate parameters and initial conditions we imposed
the artificial noise on the number of infected $I(t)$ instead of $\Delta I(t)$, partly because $\Delta I(t)$ is not the 
number of daily new infections:  it is the number of daily new infections minus the daily recoveries.
For the EAIHRD model in later section, the noise will be to imposed to $\Delta D(t)$ as it has a clear physical meaning: the daily
new deaths.

Consider the two sets of parameters shown in Table~\ref{tab:SEIR_SetParameters}.
The top row corresponds to characteristic values for an SEIR model, whereas the second (bottom) row is obtained from the parameter symmetry discussed in Sec.~\ref{sec:Input} and implemented in Eq.~(\ref{eq:SymmetrySEIR}), with a total constant population equal to $N=9039 =  S(0) + E(0)+I(0)+R(0)$. 
\begin{table}[hbtp]
\caption{Parameter and initial condition sets that respect local identifiability of the SEIR model
and the dual ($\sigma, \gamma$) symmetry.}
\label{tab:SEIR_SetParameters}
\begin{center}
\begin{tabular}{ccccccc} \hline \hline
$\beta$ & $\sigma$ & $\gamma$ & $S(0)$ & $E(0)$ & $I(0)$ & $R(0)$ \\ \hline
0.45 & 1/3 & 0.1 & 8485 & 500 & 50 & 4 \\
0.45 & 0.1 & 1/3 & $\simeq$ 28283 & $\simeq$ 1783 & 50 & -21077 \\ \hline \hline
\end{tabular}
\end{center}
\end{table}
We included the  $\simeq$ symbols since we round the model outcomes to integers.
The negative value of $R$ in the second parameter set is, of course,  not epidemiologically reasonable. It arises, however,  because physical bounds of the variables are ignored in the theoretical identifiability analysis. 

\begin{figure} [htbp]
\includegraphics[scale=0.30]{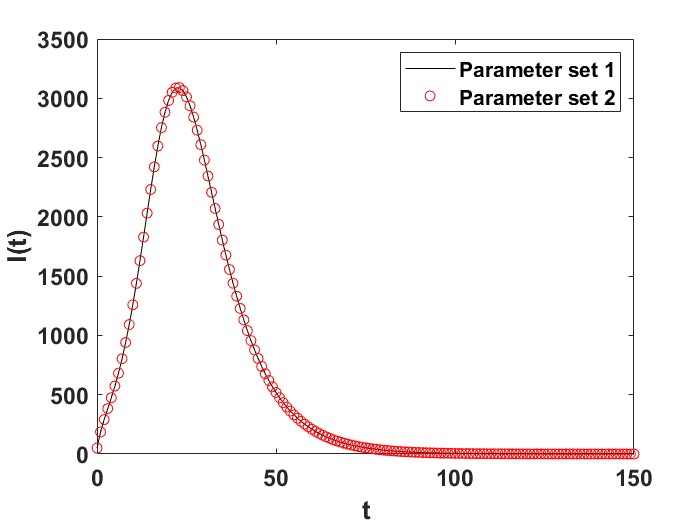} \hspace{0.2in}
\includegraphics[scale=0.30]{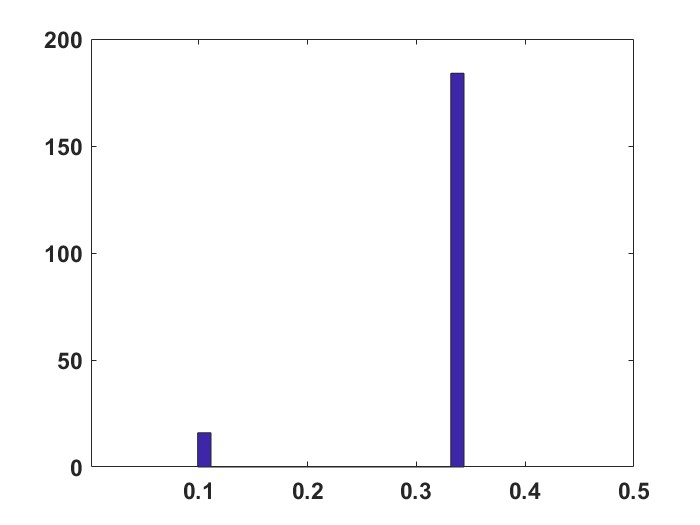}
\caption{Left: Two sets of parameters lead to the same time series of the number of infected $I(t)$, 
the reported (synthetic) data.
Right: Histogram of 200 $\sigma$ estimates. No noise was added 
to the data ($\hat{\sigma}_{\epsilon} = 0$), 
and bootstrapping was not employed. 
} 
\label{f:Is}
\end{figure}

Figure~\ref{f:Is}, left panel, shows that the calculated time series $I(t)$
obtained with  the two  parameter sets are practically identical. 
The right panel shows the 200 estimates for $\sigma$ from the DO estimator. 
We used the following intervals: 
$$
\beta \in [0.3, 0.7], \quad \gamma \in [0.001, 0.5], \quad \sigma \in [0.001, 0.5], \quad 
S(0) \in [5000, 30000], \quad E(0)\in [200, 2400].  
$$
Due to the local identifibility property of $\sigma$, more than one possible values of $\sigma$ were obtained.
The histogram may not necessarily reflect the probability of obtaining a specific $\sigma$: for each realization, the estimated $\sigma$ depends on where the initial guess is located. The ($\sigma,\gamma$) symmetry arises because $\sigma$ is locally identifiable: it can be removed by assuming that the exact $S(0)$ is given (i.e., by not treating it as an unknown to be estimated), thereby rendering the whole system globally identifiable.  

To make the observable time series more realistic we added $5\%$ noise to the data
($\hat{\sigma}_{\varepsilon}=0.05$).
In this case, we fix $S(0) =8485$ and use even larger intervals in the optimization:
\beq
\beta \in [0.01, 1.5], \quad \gamma \in [0.001, 1], \quad \sigma \in [0.001, 1], \quad  E(0)\in [10, 4000].
\label{eq:LargeBounds}
\eeq
The wider intervals mimic the common cases where everything is globally identifiable, but 
limited information  is available on the upper and lower bounds
of the intervals within which the parameters may lie.  We used
bootstrapping with $5\%$ error ($\sigma_{\epsilon}=0.05$) for the point estimators.  For the Bayesian estimator we used 5000 points (starting from the end) in the MCMC chain (with skipping ratio $10$) to predict and calculate the statistics of the estimates. 

\begin{figure}[htbp]
 \includegraphics[scale=0.20]{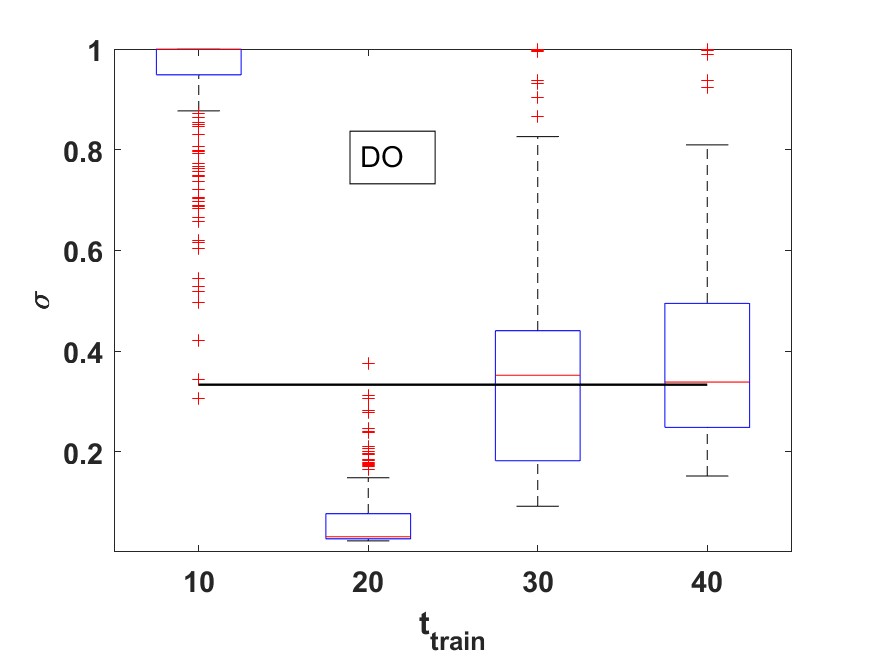} 
 \includegraphics[scale=0.20]{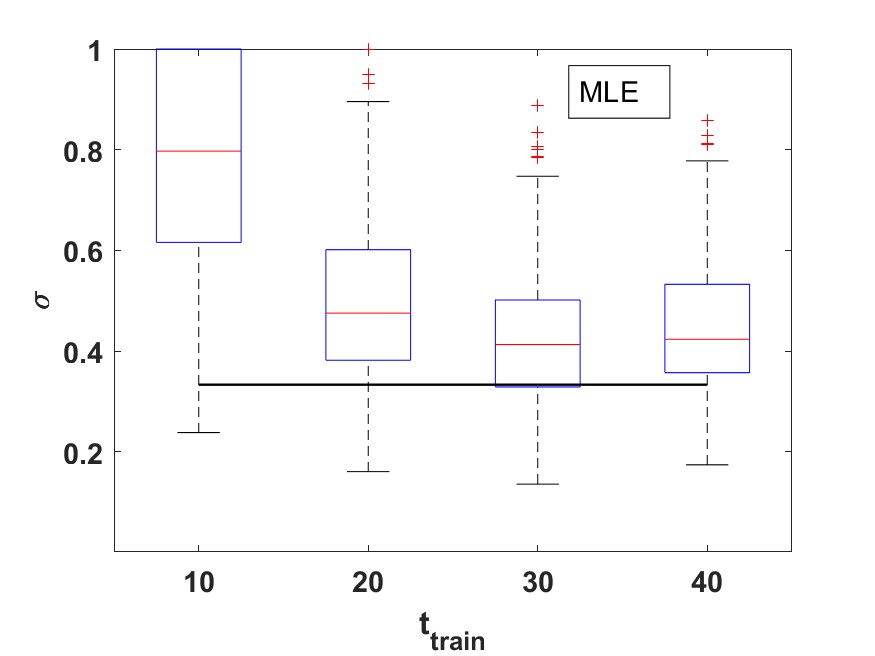} 
  \includegraphics[scale=0.20]{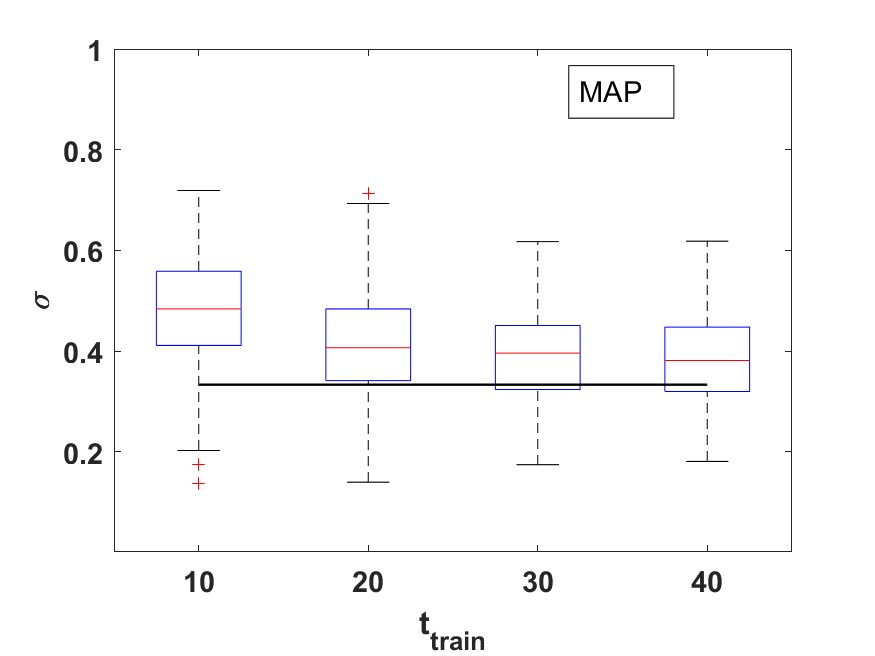} 
 \includegraphics[scale=0.20]{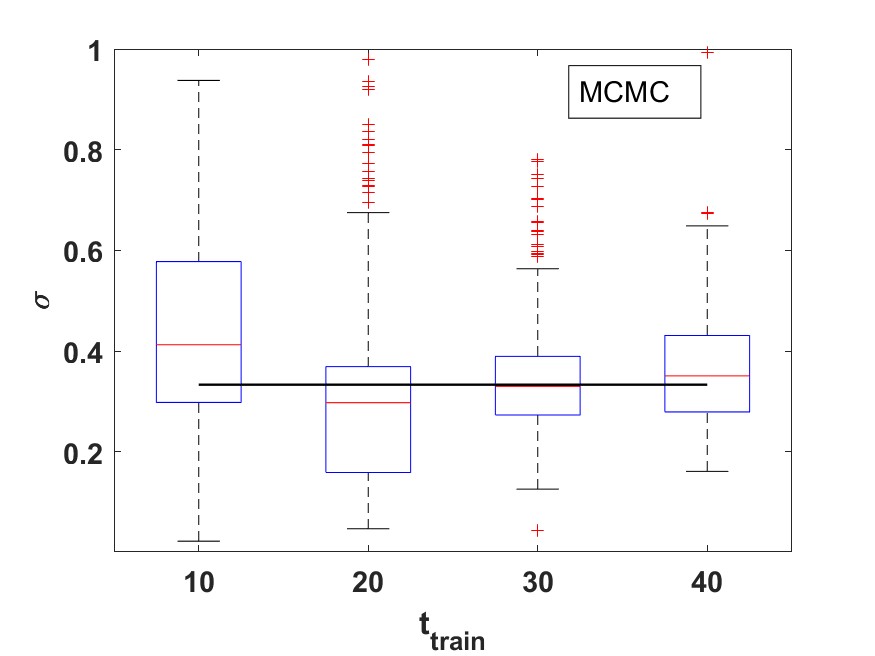} 
\caption{SEIR model.  Box plots of $\sigma$ estimates as a function of
the algorithm training period $\tTrain = 10,20,30,40$. 
Top row: DO, MLE. Bottom row: MAP, MCMC. Bootstrapping noise level $\sigma_{\varepsilon}= 0.05$ (DO, MLE, MAP).
Fixed  $S(0)=8485$. Horizontal line: exact $\sigma = 1/3$.
Data noise level $\hat{\sigma}_{\varepsilon} = 0.05$. 
Large parameter bounds,  Eq.~(\ref{eq:LargeBounds}).
}
 \label{f:FixS0_boxplots_sigma}  \end{figure}

\begin{figure}[htbp]
 \includegraphics[scale=0.125]{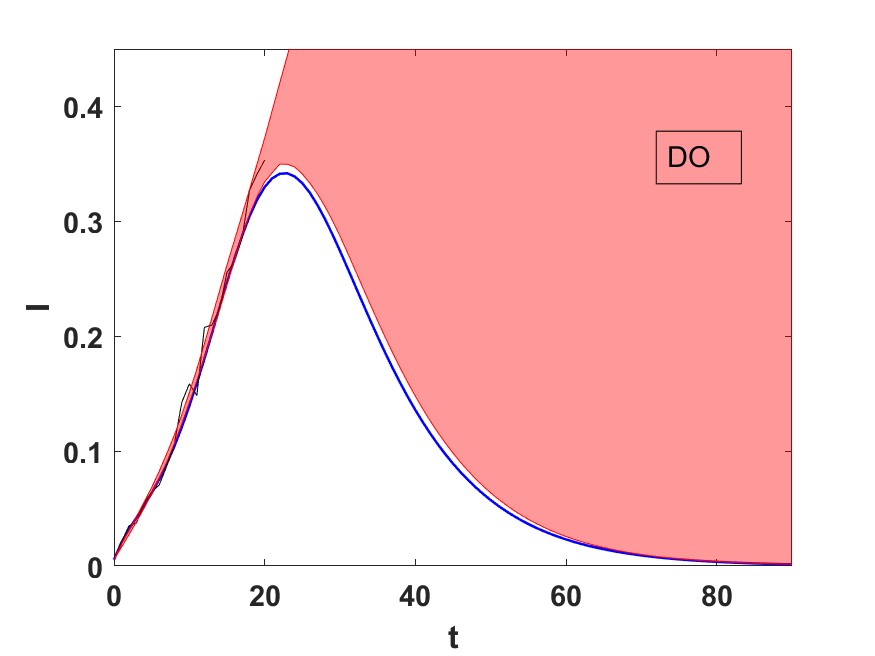} 
 \includegraphics[scale=0.125]{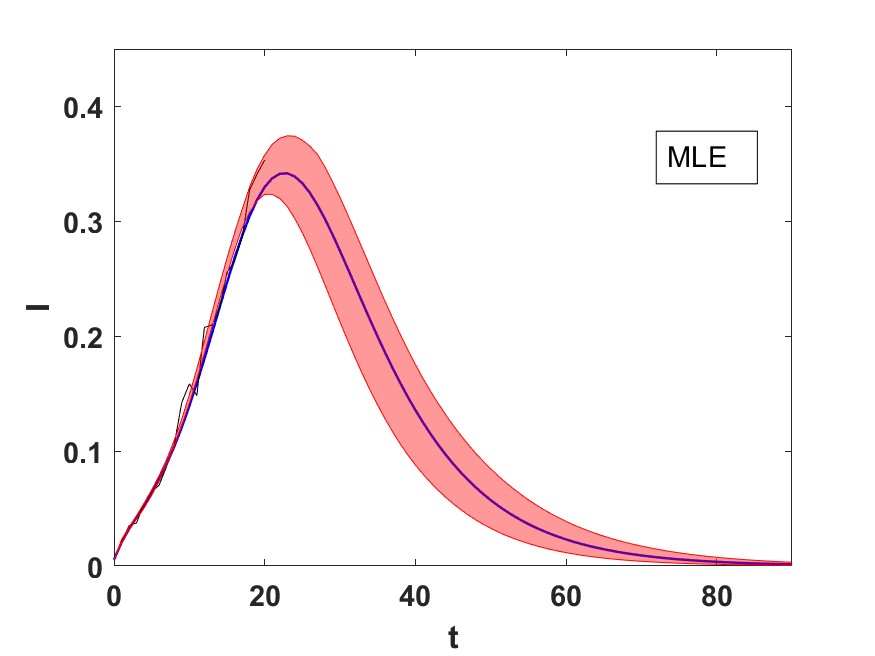} 
 \includegraphics[scale=0.125]{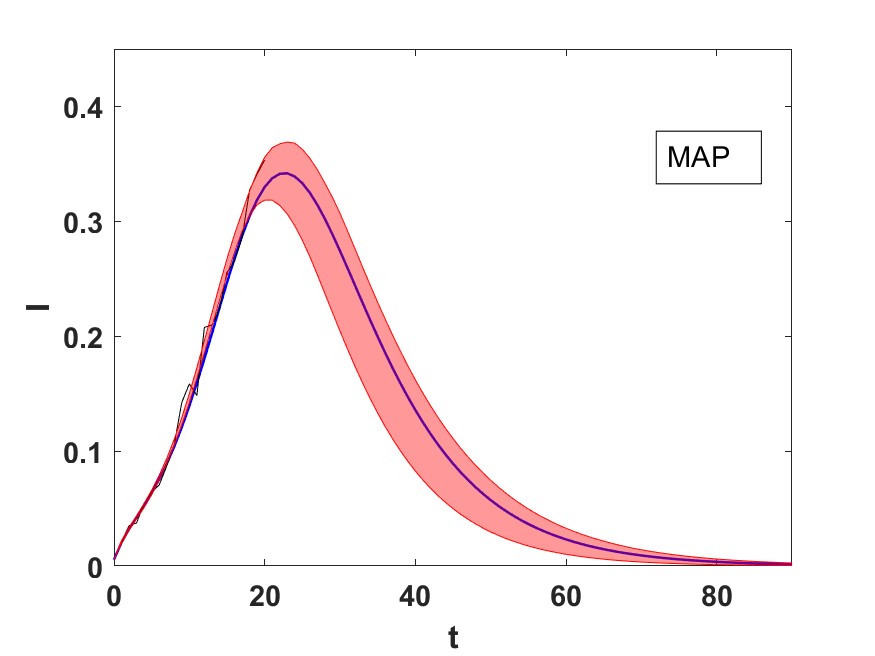}  
 \includegraphics[scale=0.125]{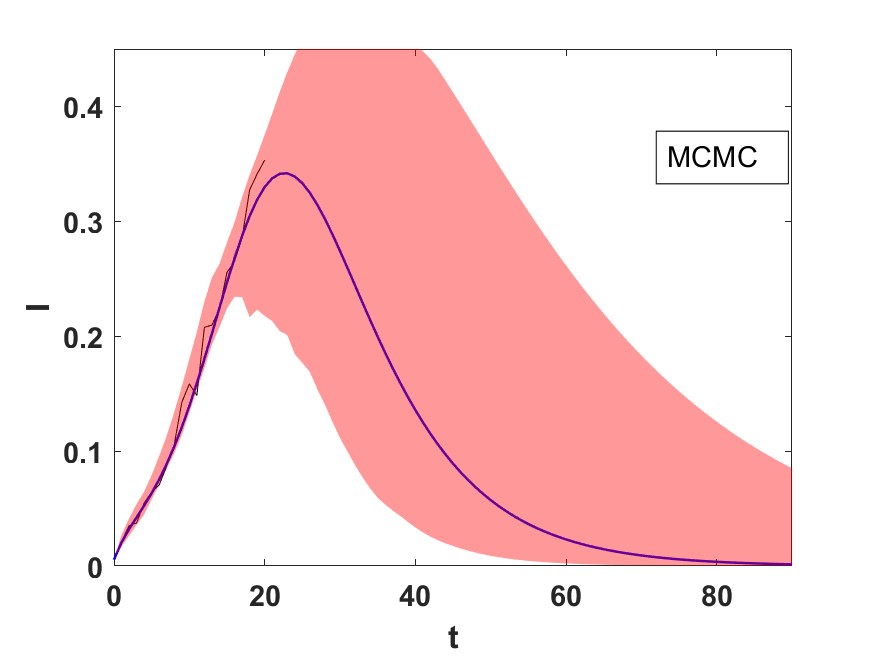}  
 \includegraphics[scale=0.125]{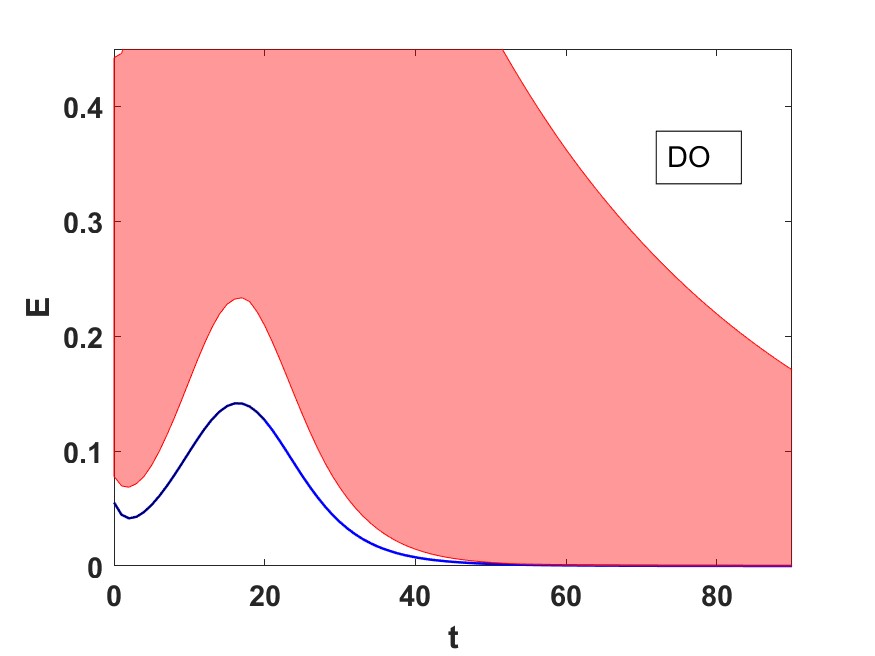} 
 \includegraphics[scale=0.125]{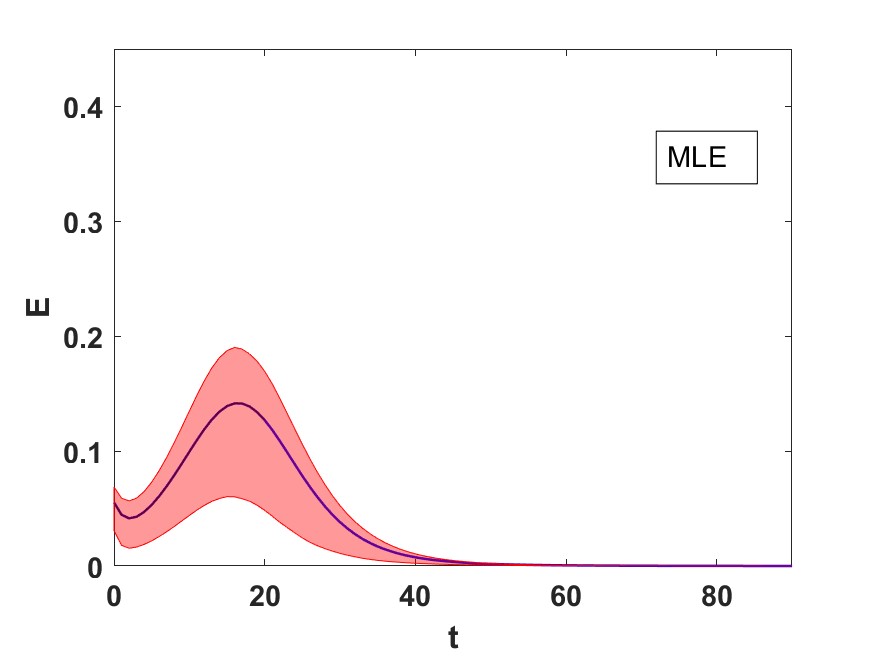} 
 \includegraphics[scale=0.125]{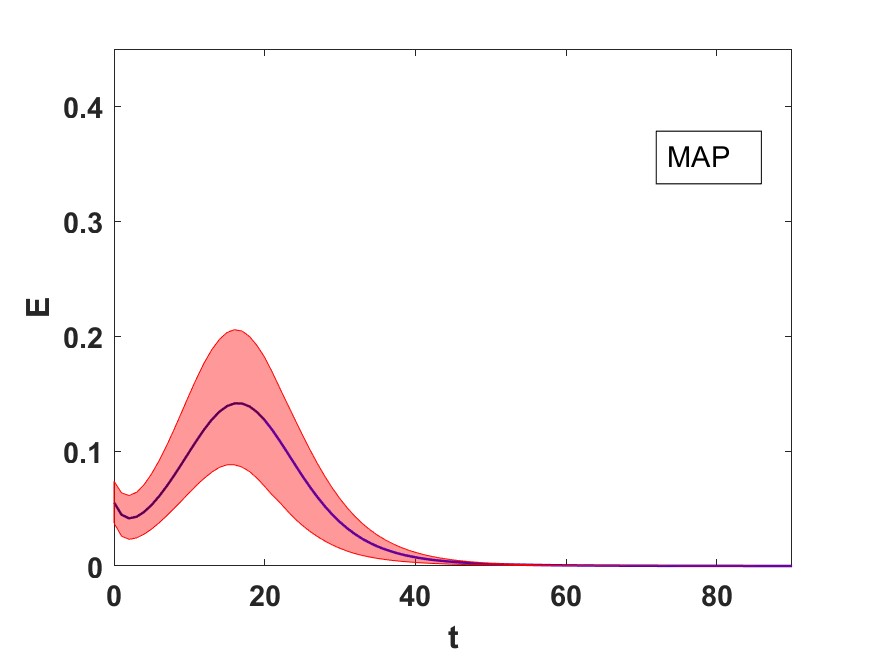} 
 \includegraphics[scale=0.125]{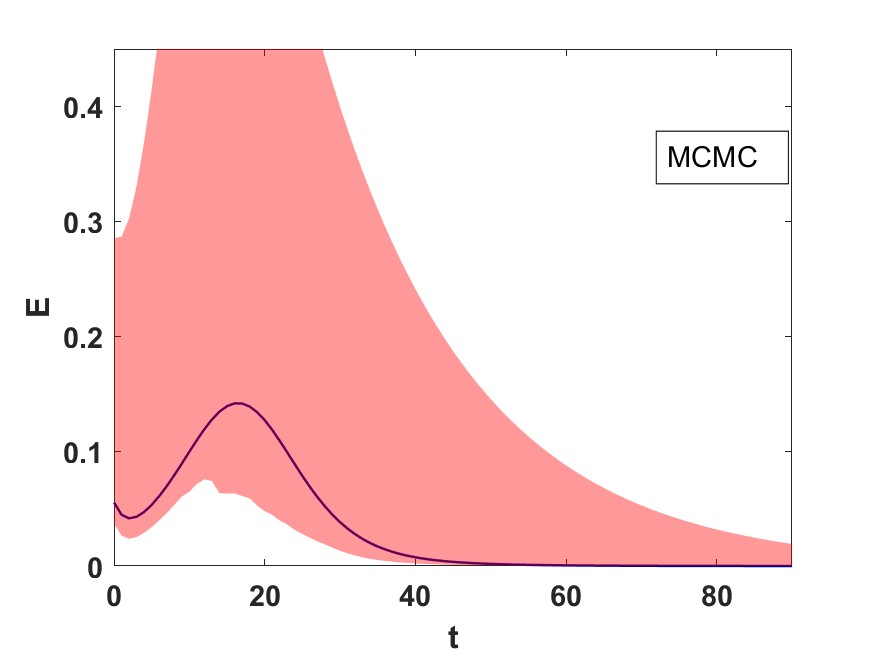}  
 \caption{SEIR model. Predicted number of infected $I(t)$ (top row) and exposed $E(t)$ (bottom row)
expressed as a fraction of the total population.
Solid $I(t)$ or $E(t)$ curves:  reported data, see main text.
Non-smooth curve for $I(t)$: polluted data used to estimate the parameters ($\hat{\sigma}_{\varepsilon} = 0.05$).
Shaded region: $5-95\%$ percentile of predictions.
Subpanels grouped in vertical pairs: first left pair represents DO predictions,
second vertical pair  represents MLE predictions,  third vertical pair yields the MAP predictions,
and  fourth vertical pair concerns the MCMC 
predictions. Bootstrapping noise level $\sigma_{\varepsilon}= 0.05$ (DO, MLE, MAP).
Fixed  $S(0)=8485$. Exact $\sigma = 1/3$.
Large parameter bounds, Eq.~(\ref{eq:LargeBounds}).
Time to train the optimization algorithms $\tTrain = 20$.
}
\label{f:FixS0_IE_Daily5DailyBS5_t20}  \end{figure}

\begin{figure}[htbp]
 \includegraphics[scale=0.125]{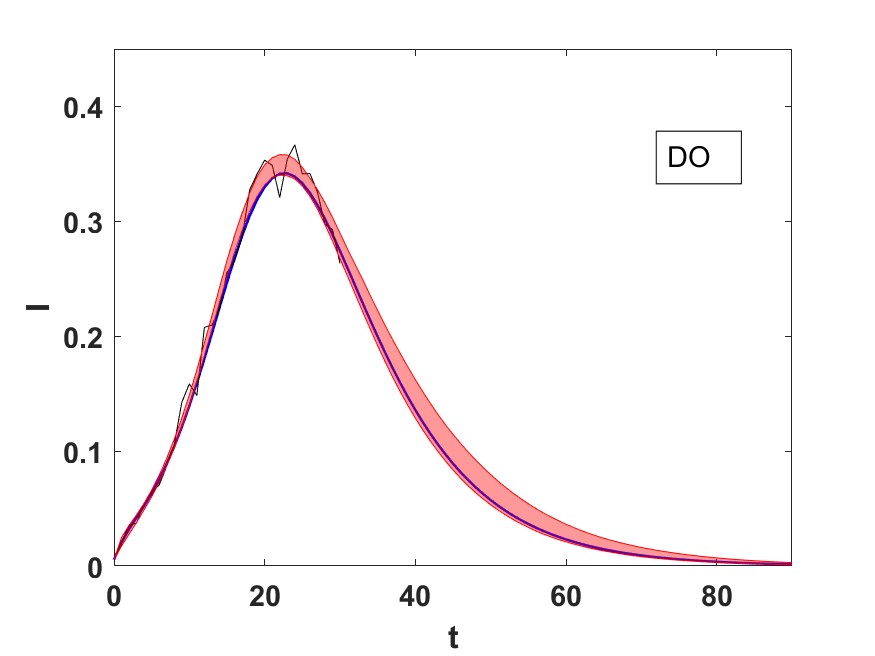} 
\includegraphics[scale=0.125]{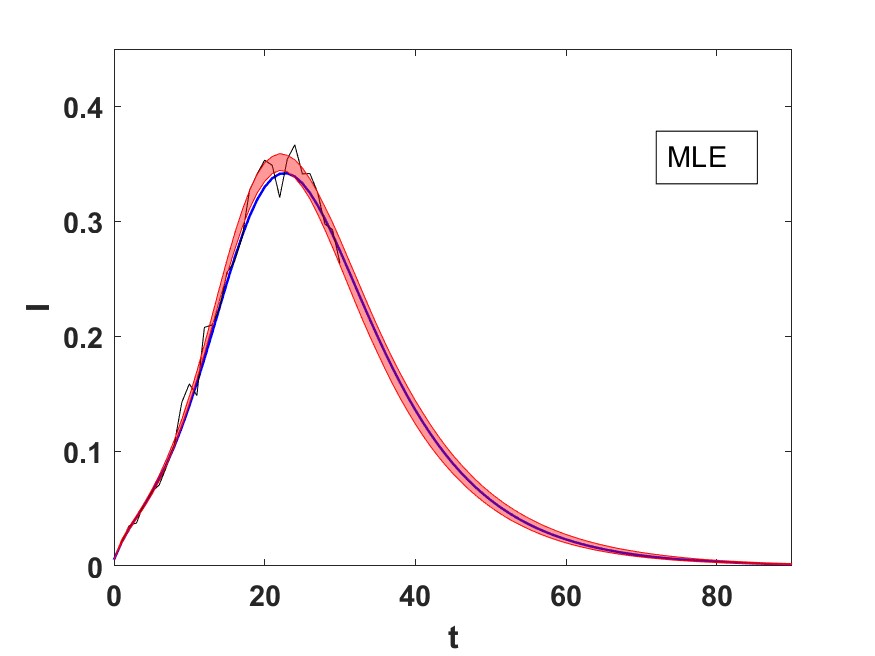}
 \includegraphics[scale=0.125]{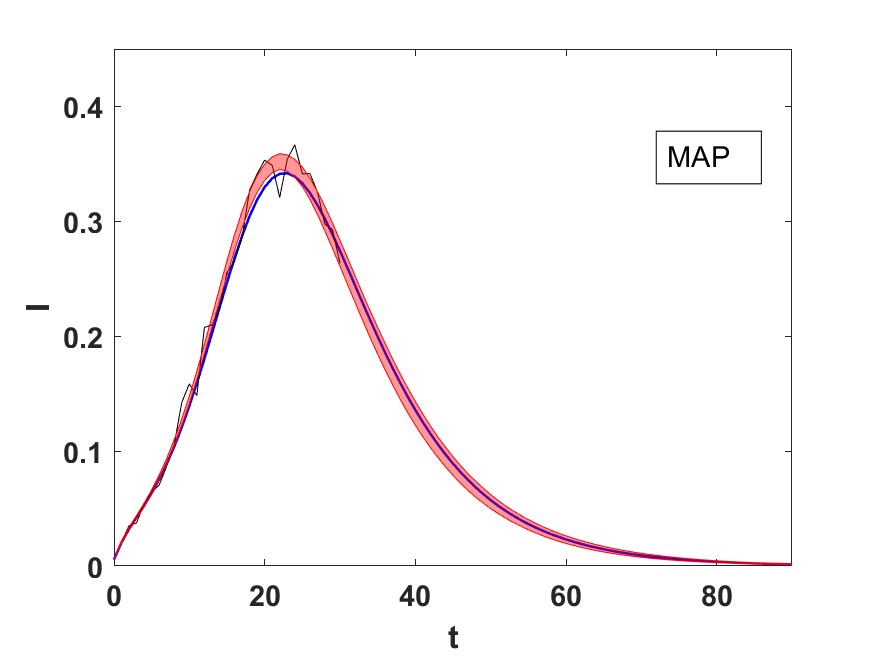} 
 \includegraphics[scale=0.125]{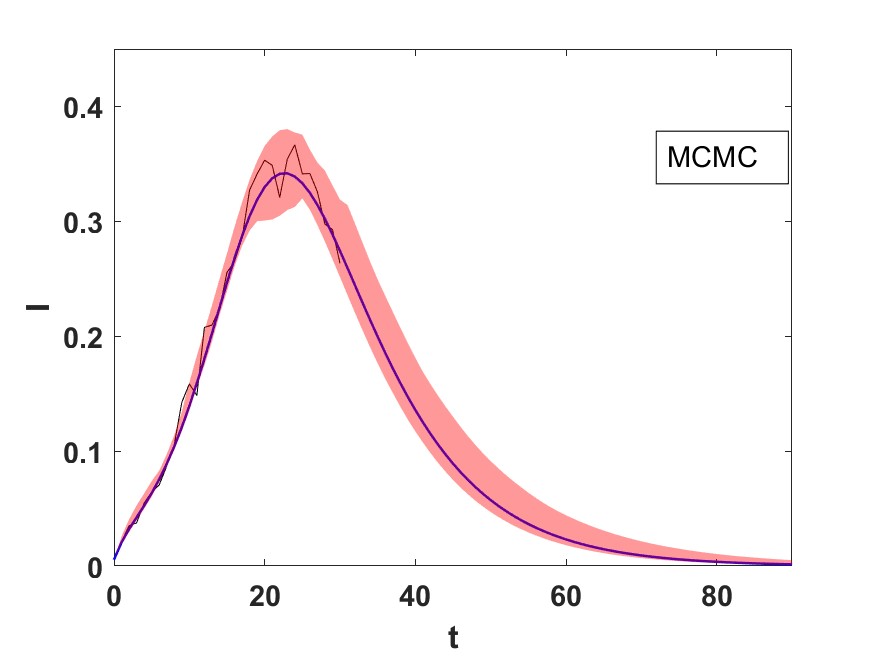}
 \includegraphics[scale=0.125]{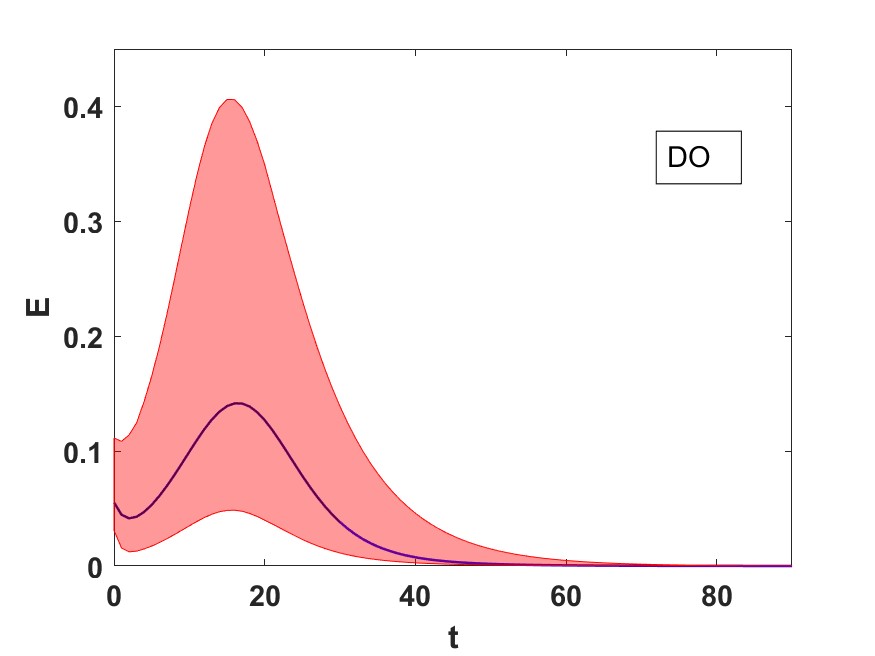}  
   \includegraphics[scale=0.125]{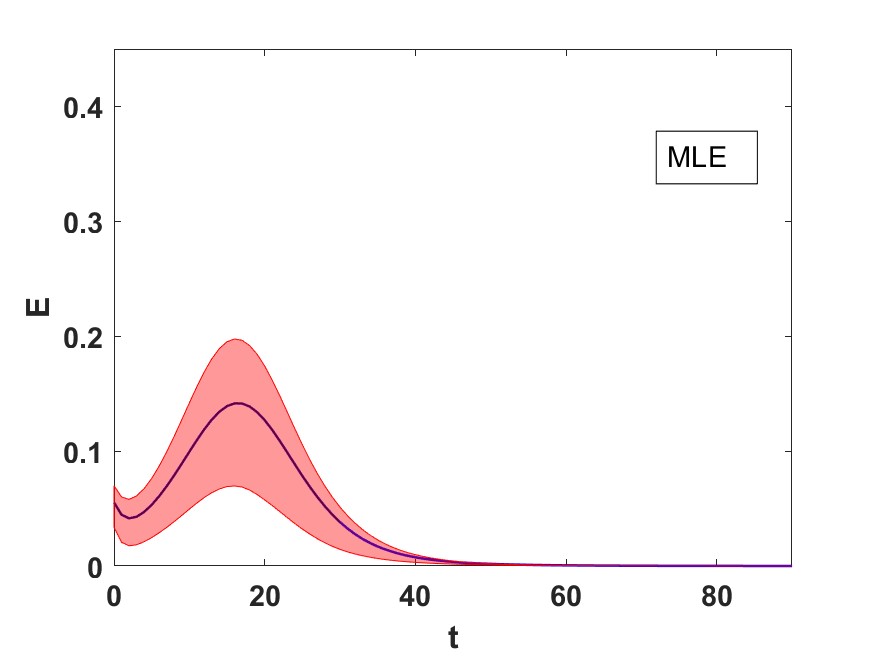}  
 \includegraphics[scale=0.125]{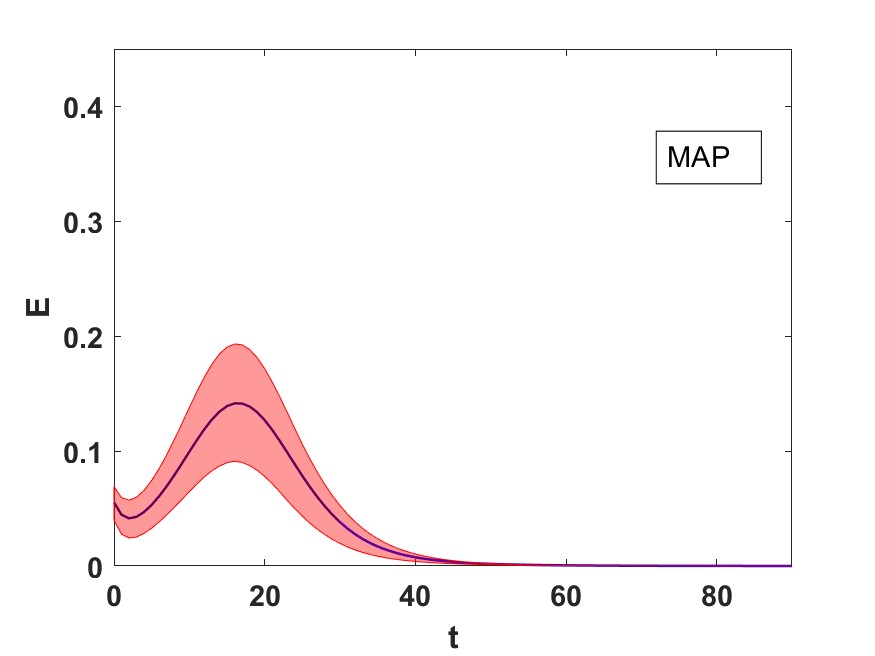} 
 \includegraphics[scale=0.125]{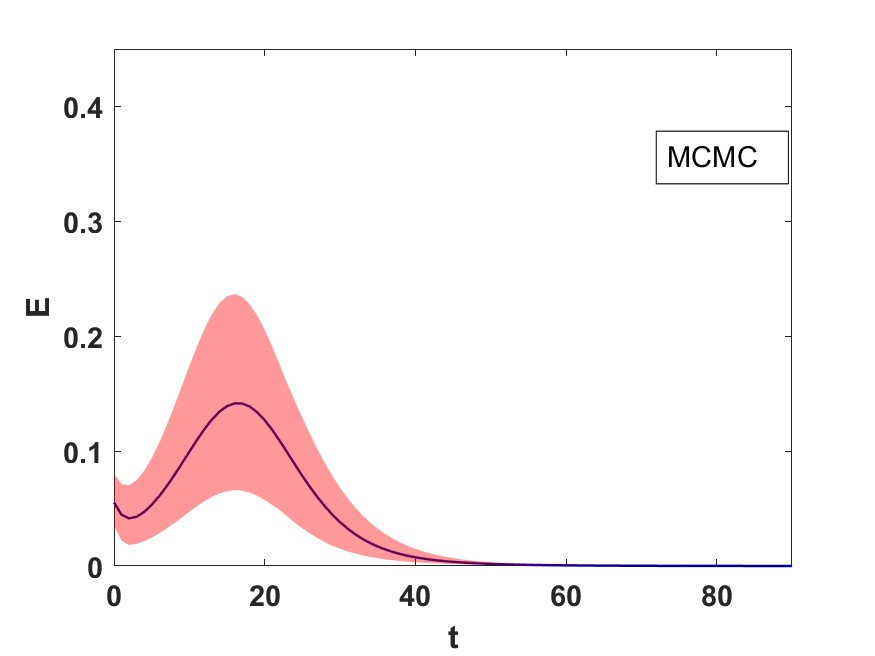}  
\caption{As in Fig.~\ref{f:FixS0_IE_Daily5DailyBS5_t20}, except that the
algorithm training period was $\tTrain = 30$.}
 \label{f:FixS0_IE_Daily5DailyBS5_t30}  \end{figure}

Figure~\ref{f:FixS0_boxplots_sigma} shows estimates of $\sigma$ according to the
four estimators in box plots as a function of $\tTrain$, the time interval used
to train the estimator's optimization algorithm.
For estimator DO, dynamical compensation~\cite{sauer2021} is apparent - there is a dramatic change in terms of the quality of the estimates when the training time changes from
$\tTrain =20$ (pre-peak) to $\tTrain=30$ (post-peak).
The change is more gradual for the other estimators. 
When only pre-peak data are used ($\tTrain=20$), the MLE and MAP estimators give very good predictions for both $I$ and $E$ 
(Fig.~\ref{f:FixS0_IE_Daily5DailyBS5_t20}), even though their $\sigma$ estimates may not be considered acceptable: the exact value is outside the $25-75\%$ percentile of the estimates (cf Fig.~\ref{f:FixS0_boxplots_sigma}).
Note that not the whole region is shown for DO or MCMC in Fig.~\ref{f:FixS0_IE_Daily5DailyBS5_t20} since we want to compare them
with the results when $\tTrain=30$ (shown in Fig.~\ref{f:FixS0_IE_Daily5DailyBS5_t30}).
As discussed above,
in addition to model predictions (with the associated errors) for the
observable $I(t)$ we also show predictions for a non-observable variable, which we chose to be
the number of exposed individuals $E(t)$.  We note that for the MLE and MAP estimators
model predictions for both the observable and the non-observable are rather good,
even though the parameter estimates are not.  This may be considered as an indication
of the sloppiness of the system, and it is a model-dependent feature.

\begin{figure}[htbp]
\includegraphics[scale=0.175]{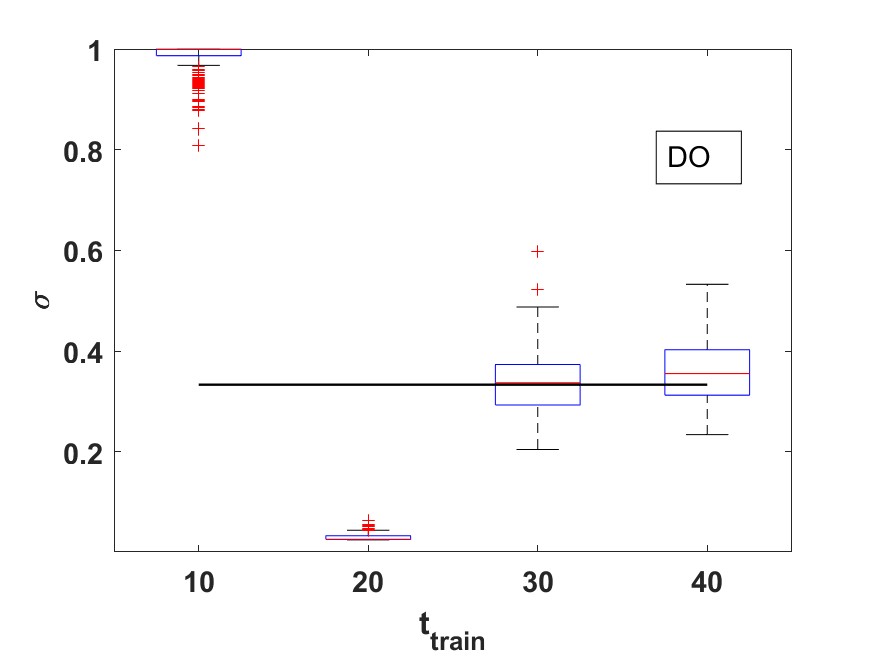} 
 \includegraphics[scale=0.175]{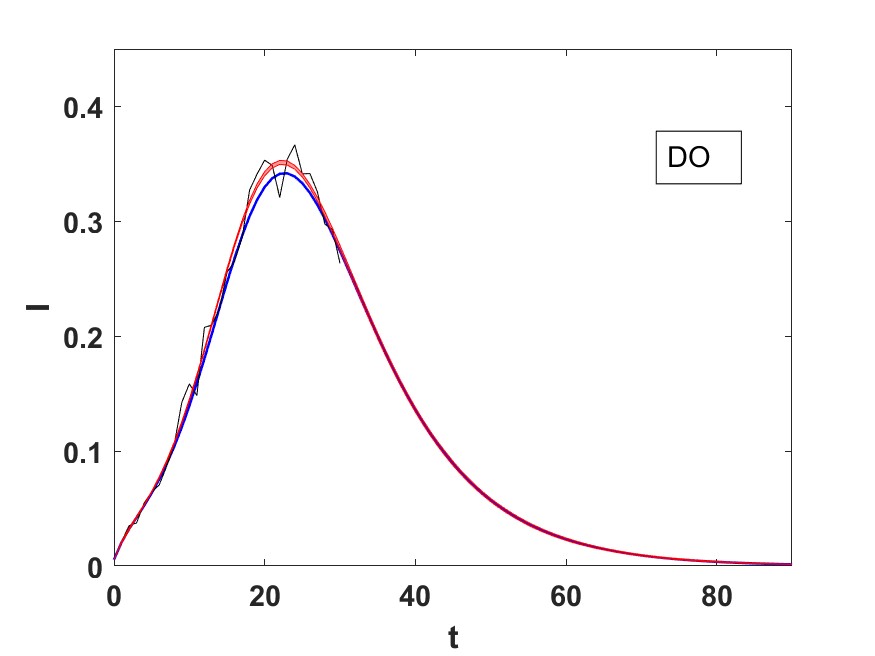} 
 \includegraphics[scale=0.175]{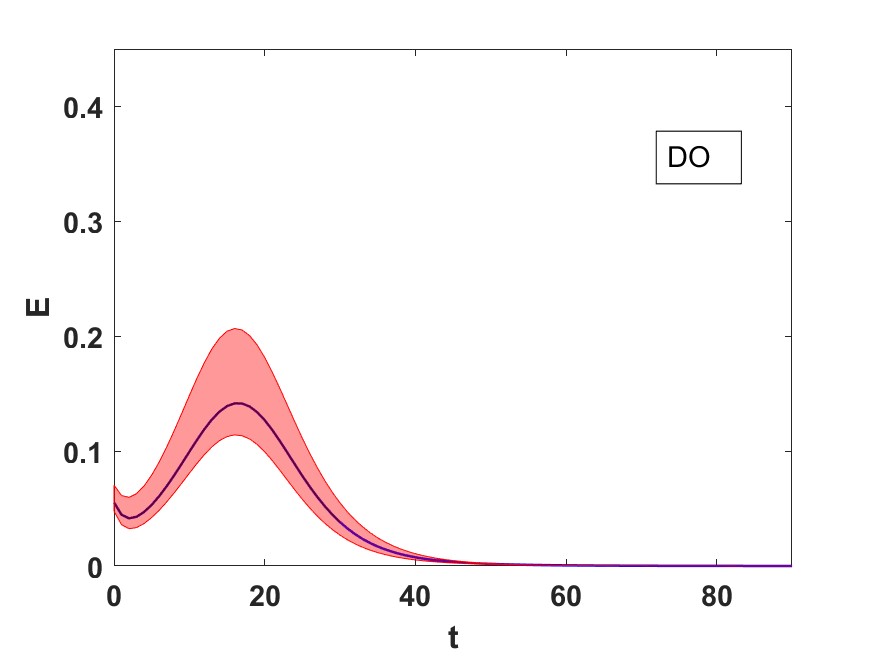} 
 \includegraphics[scale=0.175]{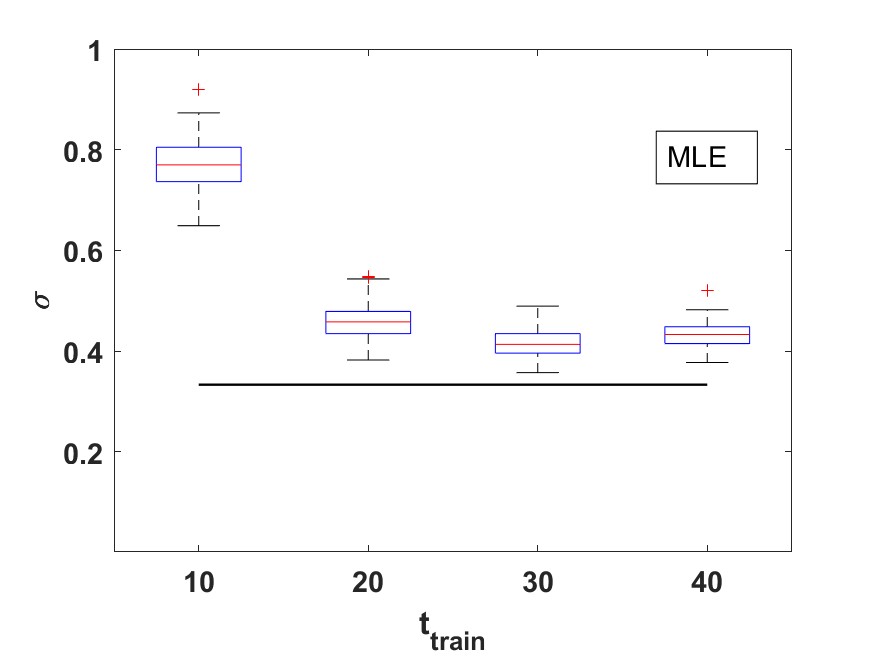} 
 \includegraphics[scale=0.175]{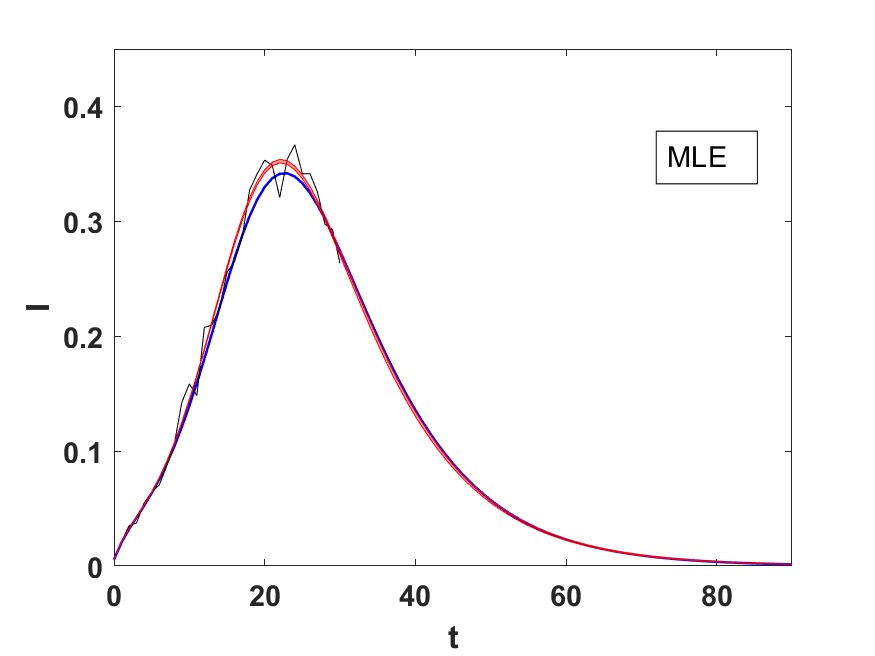}
 \includegraphics[scale=0.175]{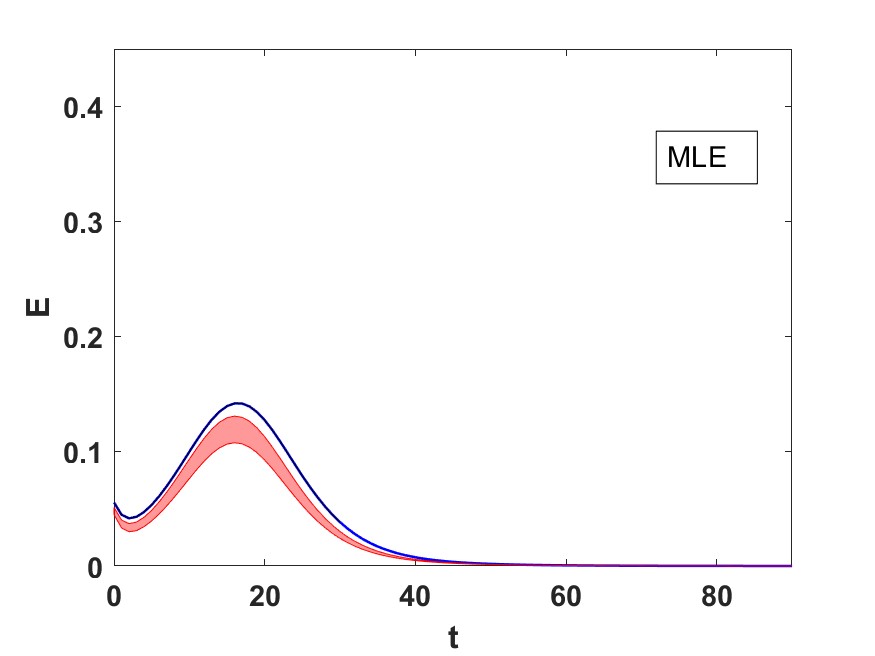} 
 \includegraphics[scale=0.175]{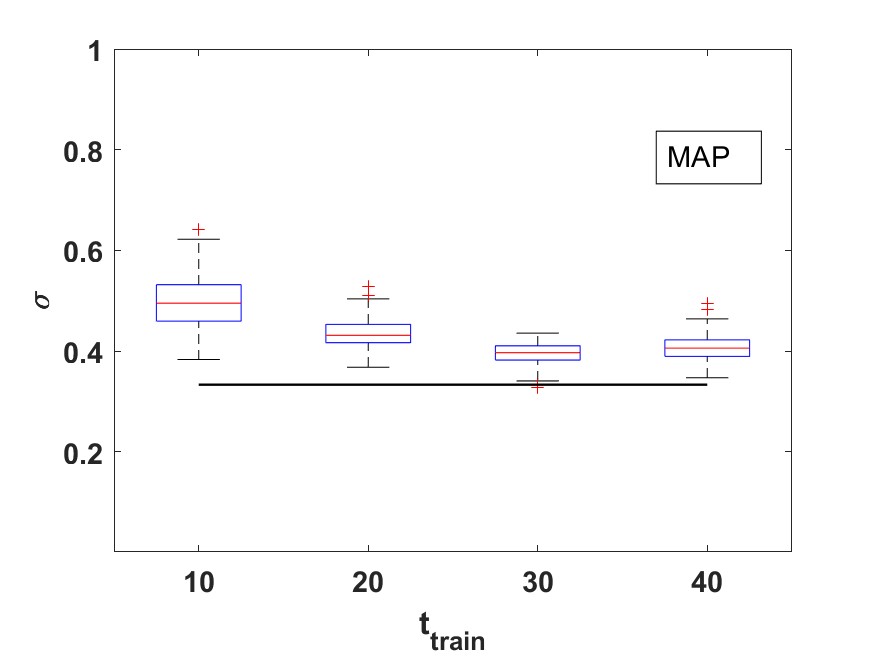}  
 \includegraphics[scale=0.175]{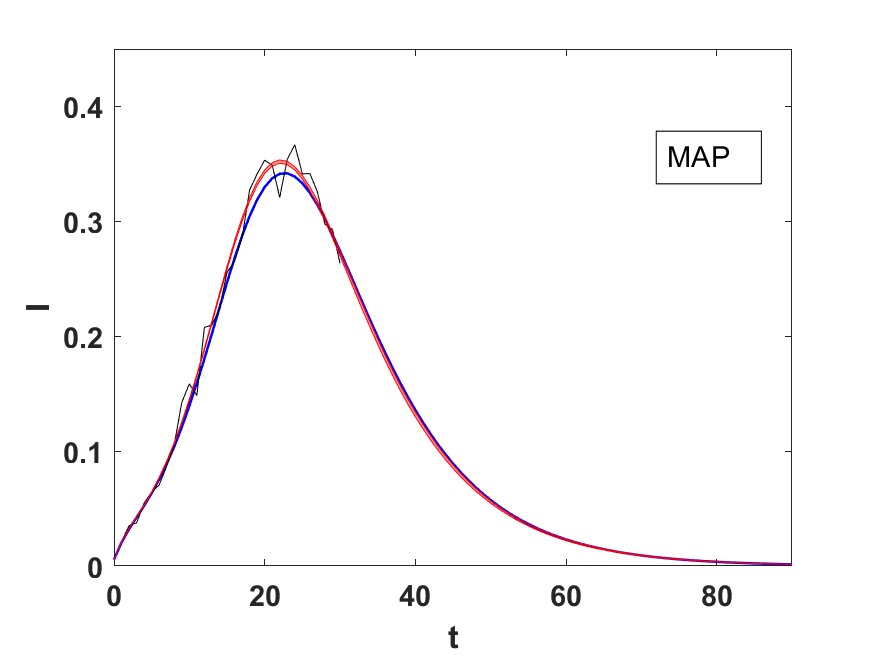}
 \includegraphics[scale=0.175]{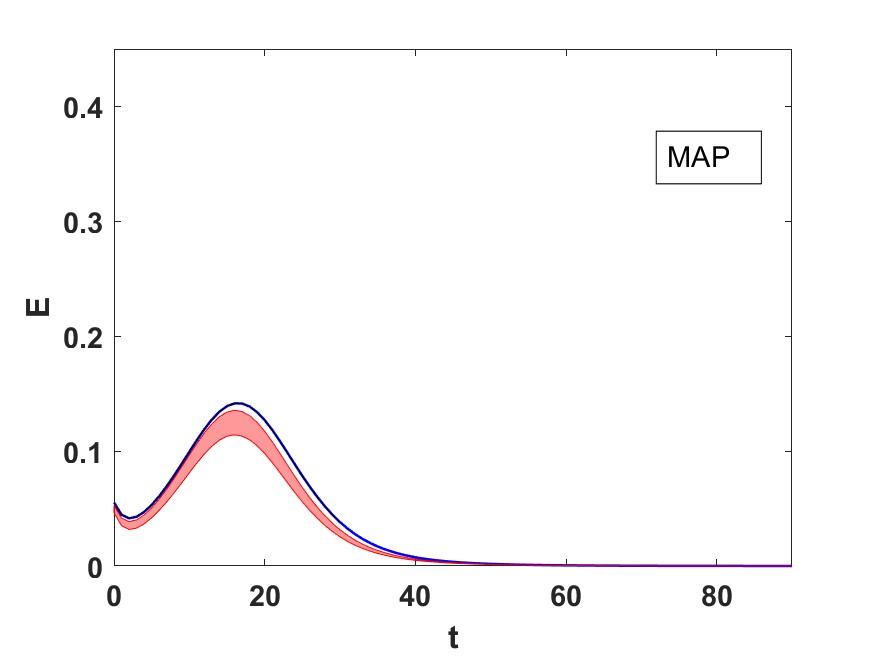}  \hfill
 \caption{SEIR model. Point estimator predictions: DO (top row),
 MLE (middle row), and MAP (bottom row). 
 Note that, in contrast to the predictions shown in 
 Figs. ~\ref{f:FixS0_boxplots_sigma}, \ref{f:FixS0_IE_Daily5DailyBS5_t20}, and
 \ref{f:FixS0_IE_Daily5DailyBS5_t30}, 
the bootstrapping noise level is lower $\sigma_{\varepsilon} = 0.01$.
Left column: Box plots of $\sigma$ estimates. Exact $\sigma = 1/3$; Middle column: 
predicted number of infected $I(t)$; Right column: predicted number of exposed $E(t)$.
As in the Fig.~\ref{f:FixS0_IE_Daily5DailyBS5_t20} caption, 
solid $I(t)$ or $E(t)$ curves reproduce the (synthetic) reported data,
whereas the jagged $I(t)$ curve presents the polluted data 
used to estimate the parameters ($\hat{\sigma}_{\varepsilon} = 0.05$).
Shaded region: $5-95\%$ percentile of predictions.
Fixed  $S(0)$,  large parameter intervals, Eq.~(\ref{eq:LargeBounds}),
and algorithm training period $\tTrain = 30$.}
 \label{f:FixS0_IE_siga_t30} \end{figure}

For a training period that contains the peak in the data, and
specifically for $\tTrain=30$, cf. Fig. ~\ref{f:FixS0_IE_Daily5DailyBS5_t30}, estimators DO, MLE, MAP and MCMC all give acceptable results. 
However, the three point estimators (DO, MLE, MAP) rely on an empirical choice of the noise level $\sigma_{\varepsilon}$ in the bootstrapping. 
 The quality of the estimates varies greatly with the choice of $\sigma_{\varepsilon}$.
For example, Fig.~\ref{f:FixS0_IE_siga_t30} presents  results for a small value $\sigma_{\varepsilon}=0.01$
that should be contrasted with those presented in the
box plots of Fig.~\ref{f:FixS0_boxplots_sigma} or the
predictions of $I(t)$ and $E(t)$ shown in Fig. ~\ref{f:FixS0_IE_Daily5DailyBS5_t30}.
Clearly, with a smaller value  $\sigma_{\varepsilon} = 0.01$ the DO estimator gives a better estimate for $\sigma$ 
and a much better prediction for $E$. 
However, estimators MLE and MAP now yield unacceptable $\sigma$ estimates.

Lastly, we consider a more realistic test case where we also
address the importance of initial conditions, specifically of $S(0)$.
We investigate possible differences between 
specifying an exact value versus specifying a narrow range. As argued
in Sec.~\ref{sec:Input}, the SEIR model can be made globally identifiable either by
fixing $S(0)$, or limiting it to a relative small interval  such that a unique $S(0)$,  and
subsequently a unique set of parameters, exists.
We have shown that the global identifiability of the model ensured by fixing $S(0)$
to its exact value indeed leads to satisfactory results for all the estimators, although DO, MLE, and MAP may need
an appropriate bootstrapping noise level $\sigma_{\varepsilon}$ (=0.05 in the above results).
We now ask the question how accurately are the estimated parameters when
a small $S(0)$ interval is enforced. 
Identifiability analysis suggests that the system continues to be globally identifiable.
We not only chose a relatively small interval for $S(0)$, but also for $E(0)$ and the other parameters:
\beq
\beta \in [0.3, 0.7], \quad \gamma \in [0.001, 0.5], \quad \sigma \in [0.001, 0.5], \quad 
S(0) \in [5000, 9000], \quad E(0)\in [200, 2400].
\label{eq:SmallBoundS0}
\eeq
\begin{figure}[htbp]
\includegraphics[scale=0.30]{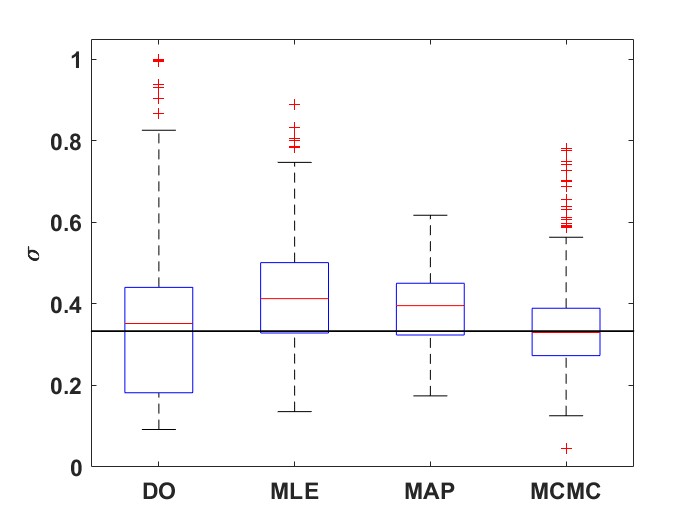} 
\includegraphics[scale=0.30]{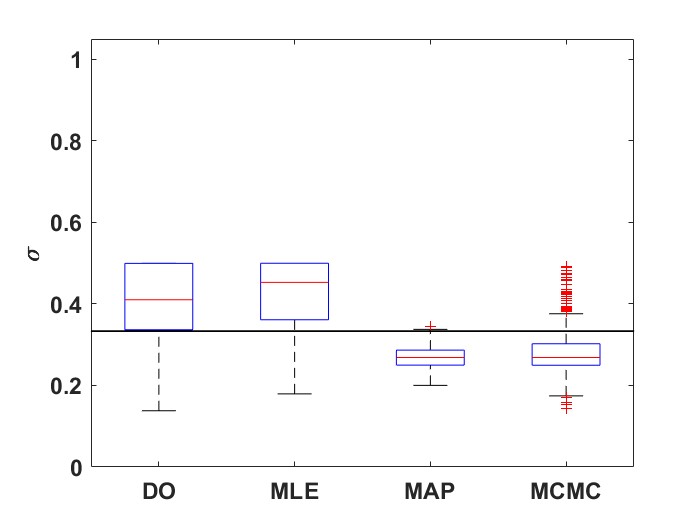} 
\caption{SEIR model: Estimates of $\sigma$ as a function of constraints
on the initial condition $S(0)$.  Left panel: $S(0)$ fixed to the exact value.
Right panel: $S(0)$ bounded to lie with a small interval, cf. Eq.~(\ref{eq:SmallBoundS0}).  Estimators
used: DO, MLE, MAP, and Bayesian MCMC. 
Noise of synthetic data $\hat{\sigma}_{\varepsilon} = 0.05$, bootstrapping noise
$\sigma_{\varepsilon} = 0.05$.  Algorithm training time interval  $\tTrain = 30$.
Horizontal line: exact value $\sigma = 1/3$.
}
\label{f:SEIR_EstimateSigmaBoxPlots} 
\end{figure}

\begin{figure}[htbp]
 \includegraphics[scale=0.125]{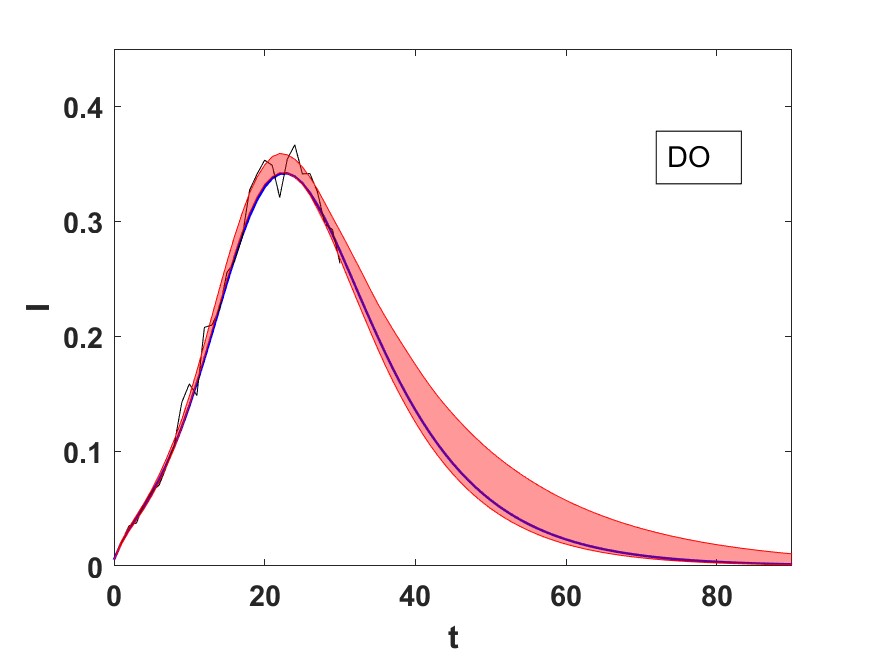} 
 \includegraphics[scale=0.125]{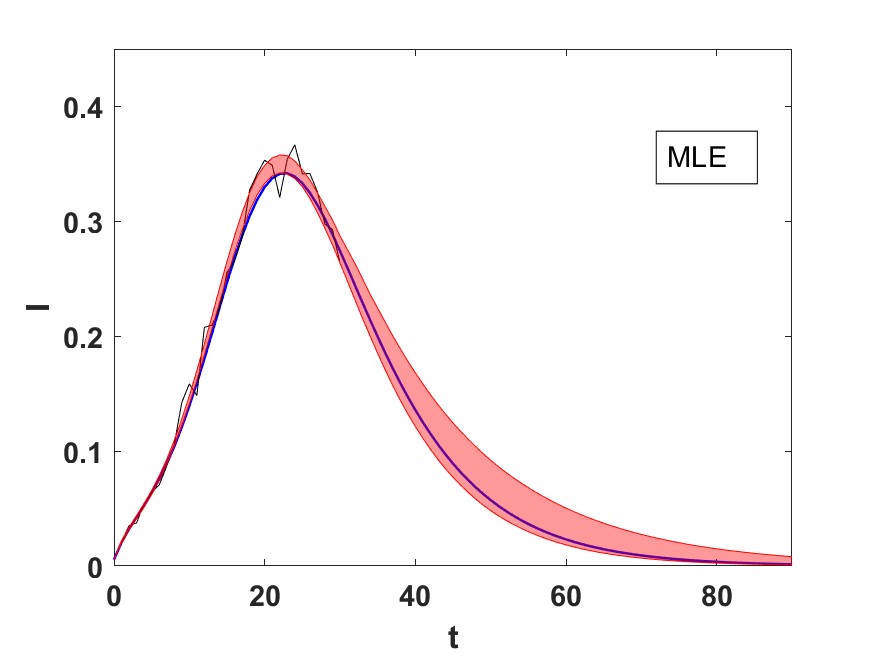} 
 \includegraphics[scale=0.125]{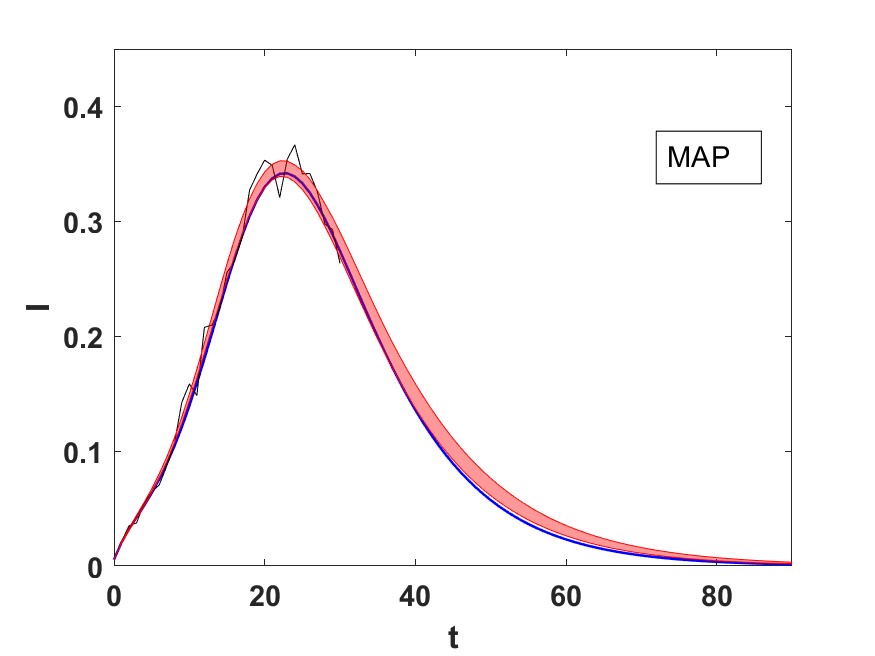}  
 \includegraphics[scale=0.125]{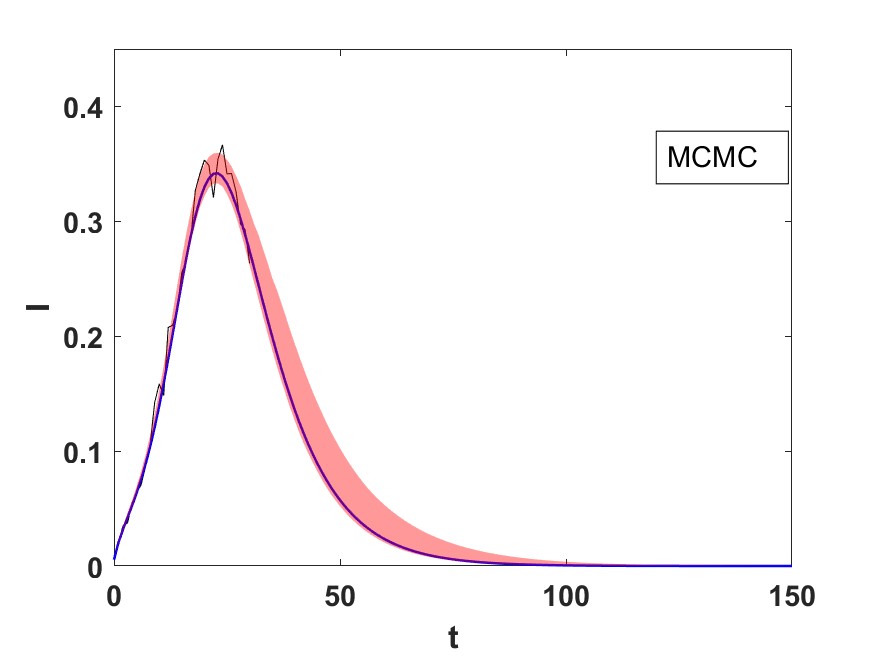}  
 \includegraphics[scale=0.125]{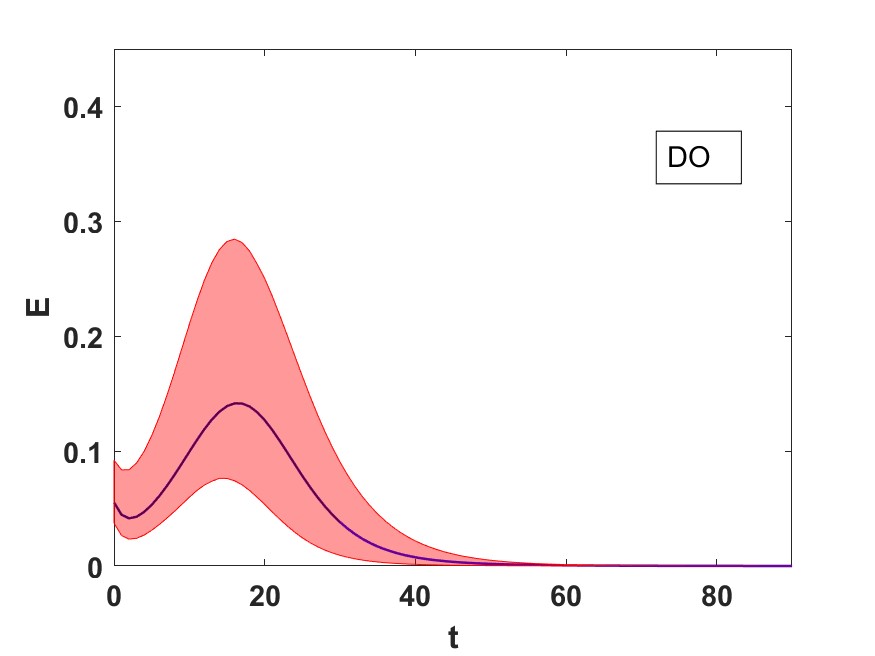} 
  \includegraphics[scale=0.125]{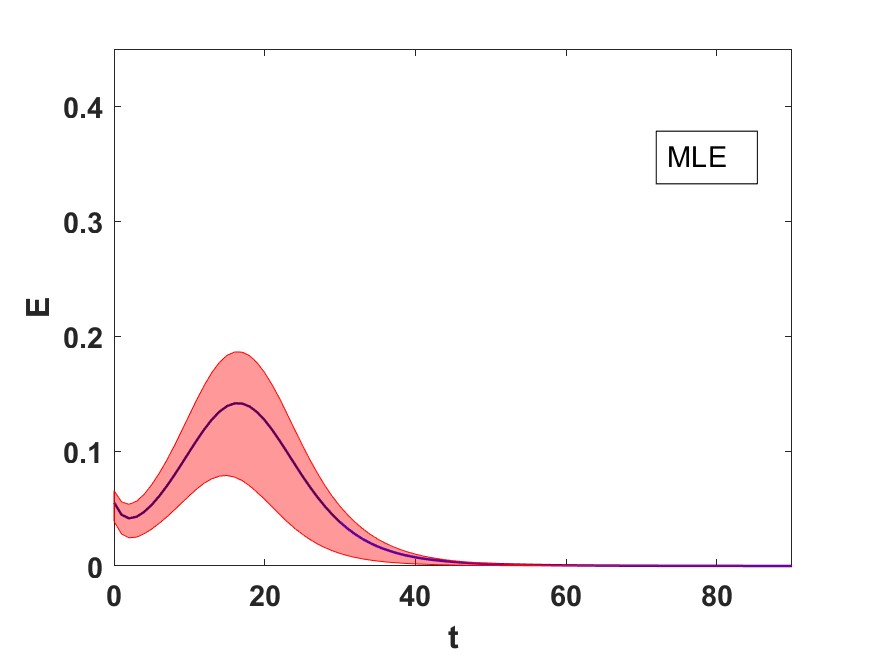}
 \includegraphics[scale=0.125]{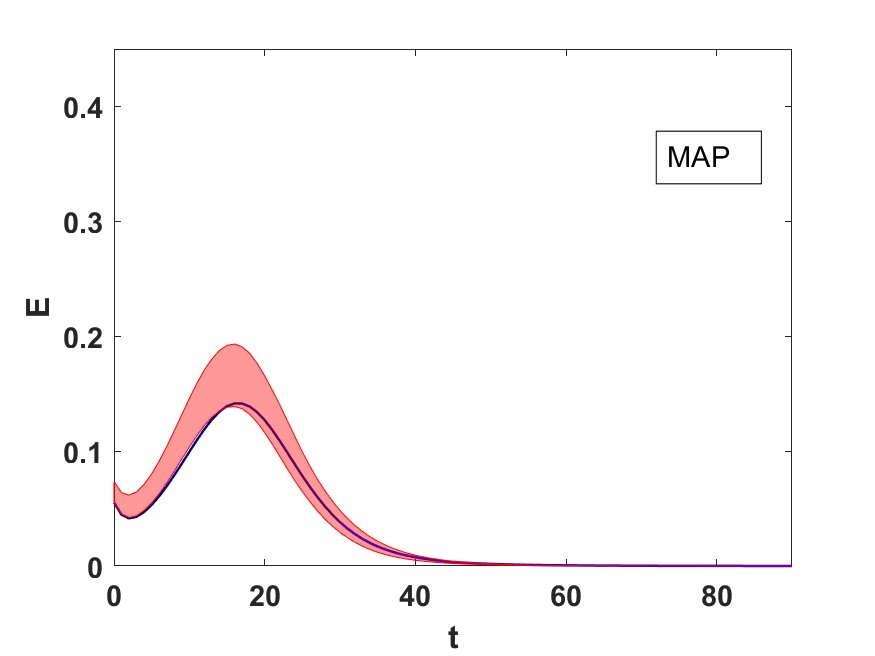} 
\includegraphics[scale=0.125]{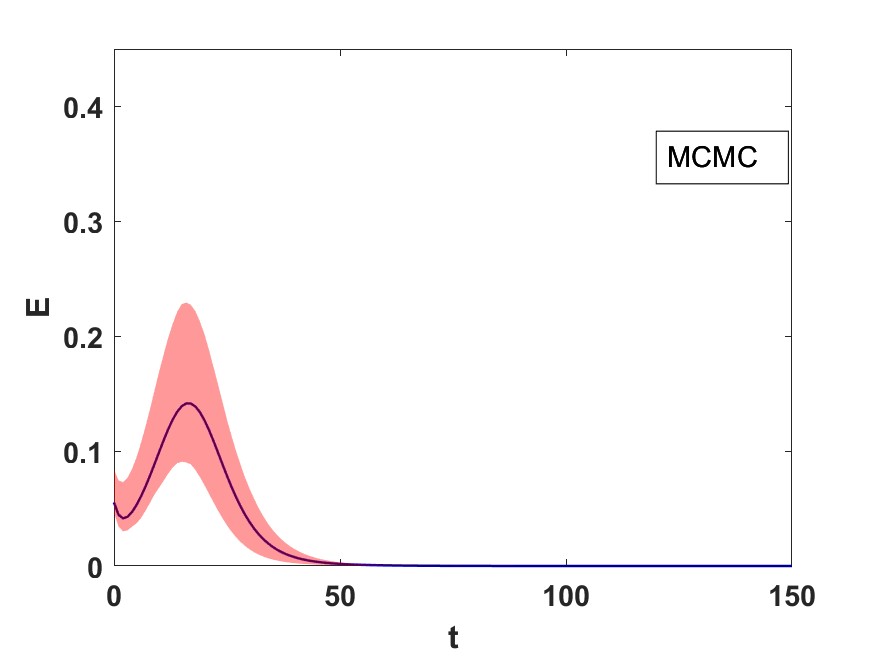} 
\caption{SEIR model.  Predicted number of infected $I(t)$ and  $E(t)$,
expressed as fraction of the total population.
Solid $I(t)$ or $E(t)$ curves:(synthetic) reported data.
Non-smooth curve for $I(t)$: polluted data used to estimate the parameters ($\hat{\sigma}_{\varepsilon} = 0.05$).
Shaded region: $5-95\%$ percentile of predictions.
Bootstrapping noise level $\sigma_{\varepsilon}= 0.05$ (DO, MLE, MAP).
Initial condition $S(0)$ restricted to a small interval, cf. Eq.(\ref{eq:SmallBoundS0}). Exact $\sigma = 1/3$.
Time to train the estimation algorithms $\tTrain = 30$.
 } \label{f:SEIR_IE_30_Daily5DailyBS5}\end{figure}

In theory, the relatively small interval for $S(0)$ should effectively eliminate a
second set of parameters that could lead to the same observable time series.
However, the parameter estimates are not as accurate as those obtained by fixing $S(0)$
the case that ensures global identifiability.
We only show the estimates of $\sigma$ in Fig.~\ref{f:SEIR_EstimateSigmaBoxPlots} as box plots,
for brevity.
Similar observations hold for the other parameters.
The left panel shows the estimates with the initial condition
of $S(0)$ fixed, whereas the right panel shows those obtained when $S(0)$ variation is bounded to a
small interval. For a variable $S(0)$, in the right panel,
the $25-75\%$ percentile of every estimator
does not even include the exact value for $\sigma$. 
The estimates are worse  
even though tighter bounds were enforced on all parameters and initial conditions, except $S(0)$. 
Figure~\ref{f:SEIR_IE_30_Daily5DailyBS5} shows the corresponding predictions for
 $I(t)$ and $E(t)$  with the estimated
parameters (to be compared to Fig.~\ref{f:FixS0_IE_Daily5DailyBS5_t30}).
Even though the parameter estimates obtained by fixing $S(0)$ and 
optimizing
are considerably different, the predictions for the observable $I(t)$ and the non-observable $E(t)$
do not differ greatly. This represents another sign of the sloppiness of the SEIR model.

We showed that these two different ways to achieve global identifiability
may lead to
different parameter estimates. We believe the reason is as follows.
At the structural level, the model becomes intrinsically globally identifiable when we fix $S(0)$.
However, when $S(0)$ variations are limited to a small interval, a sharp estimate of $S(0)$ (or other initial conditions and parameters) 
cannot be guaranteed because the output $I(t)$ is rather insensitive,  namely it 
depends weakly, on the initial condition $S(0)$.
More precisely,  the system is in theory globally identifiable when there is no noise in $I(t)$. 
But when noise is present, global identifiability may not be practically
attained due to 
this lack of sensitivity of the observable on some parameters (or initial conditions).
The prescribed relatively small bounds may
not guarantee the uniqueness of parameter and initial conditions for noisy data. 

Parameter sensitivity of a fixed output,  i.e., how sensitive the
model output is to parameter variations, may be assessed by considering
the eigenvalues of the Hessian matrix, as discussed in Sec.~\ref{sec:Hessian}.
The eigengap, defined as the largest difference of the logarithms of two consecutive eigenvalues
is usually an indicator of the sensitivity of the
system~\cite{Stigter2017}. For the SEIR model  the gap between eigenvalues is
moderate (5.2 for fixed $S(0)$ and 5.72 for variable $S(0)$ constrained within a range) compared to what is observed for other models, and specifically it is
much smaller than the eigengap  in the model discussed in the following subsection. The relatively small
 eigengap renders the estimated
parameters and prediction of the SEIR model more accurate to a  certain extent.
Since only local sensitivity is computed (reflected in the eigenvalues), it is expected that there is essentially no
significant difference in the eigengap
whether $S(0)$ is fixed or not. 
For an extensive analysis of the sensitivity of the output of 
a rather complex model to parameter variations
via the eigenvalues and eigenvectors of the Hessian
cf.~\cite{Mexico}. The implications of the sensitivity analysis on model
identifiability are also discussed in~\cite{Mexico}.

\subsection{Exposed-Asymptomatic-Infected-Hospitalized-Recovered-Dead (EAIHRD)}
\label{sec:predict_fokas}

Herein, we estimate the parameters and initial conditions of the EAIHRD model~\cite{Fokas2020}, whose identifiability analysis was presented in Sec.~\ref{sec:IdentifyFokas}. We will focus on the mixed form of the model consisting of a high-order ODE for $D(t)$ and a first-order ODE 
for $\tilde{A}(t)$ derived in Sec.~\ref{sec:IdentifyFokas} and summarized in Eqs.~\eqref{eq:CSF4thA}, referred to as model 4thA, 
the name arising from the 4th order ODE for $D(t)$ and the first-order ODE for $\tilde{A}(t)$.
As in~\cite{Mexico}, the cumulative number of fatalities $D(t)$ will be used as the reported data. The noise in the data and the bootstrapping errors will be imposed on $\Delta D(t)$, which is the daily new deaths. To render the numerical tests more realistic, we opted to consider the medians of parameters and initial conditions from~\cite{Mexico} as the exact values.  Of course, the EAIHRD model 
(and hence the 4thA model) 
differs from the model presented in~\cite{Mexico} in numerous ways: in particular, the asymptomatic-susceptible transmission rate $\beta_{SA}$ and 
symptomatically infected -susceptible transmission rate $\beta_{SI}$ are time dependent in~\cite{Mexico} since public health COVID-19 policy changed with time. For that reason, we did not directly  use the COVID-19 data, $D(t)$, from~\cite{Mexico}. Moreover, it is relevant to note that the model of~\cite{Mexico} featured
a more elaborate structure including a presymptomatic population.

In addition, the initial conditions $D(0)$, $D'(0), D''(0)$ and $D'''(0)$ are required
to integrate model 4thA. The derivatives can be obtained in terms of $H(0), E(0),$ and $I(0) $ by letting $t\to0$ in Eqs.~\eqref{eq:HIE}
\begin{eqnarray*}
D'(0) &=& d H(0) , \\
D''(0) &=& hd I(0) - R_3 d H(0) , \\
D'''(0) &=& hsd E(0) - hd (R_2+R_3) I(0) + R_3^2 d H(0) .
\end{eqnarray*}
The initial conditions and the exact parameters are
\begin{eqnarray*}
F^{(1)} = 0.4721;  \; r_1 = 0.1453; \; R^{(1)}_2 = 0.1793;  \;R_3 = 0.1036;  \;C^{(1)}_1=44.1761; \; C^{(1)}_2=163.6048;  \;\alpha=0.000425; \\
\left[ D(0); \; D'(0); \; D''(0);\; D'''(0);\; \tilde{A}^{(1)}(0) \right] =  \left[ 0.0392;\; 0.0144; \; 0.0183; \;0.0058;  \;0.1702 \right] \times 10^{-6} .
\end{eqnarray*}
Note that the $D$'s and $\tilde{A}$ are normalized by the total population of Mexico $N=127\, 575\,528$. 
As shown in Sec.~\ref{sec:IdentifyFokas}, model 4thA is only locally identifiable. 
Since parameters $F+R_2$ and $F R_2$ are globally identifiable,  they induce a parameter symmetry. Consequently,
a second set of parameters  and initial conditions, denoted by the supersript $(2)$, leads to the identical time series $D(t)$,
see Table~\ref{tab:CSF_SetPar},
\begin{subequations}
\begin{eqnarray}
F^{(2)} = 0.1793;   r_1 = 0.1453; R_2^{(2)} = 0.4721;  R_3 = 0.1036;  C_1^{(2)}=4.5918;  C_2^{(2)}=203.1890;  \alpha=0.000425; \\
\left[ D(0); \; D'(0); \; D''(0);\; D'''(0);\; \tilde{A}^{(2)}(0) \right]  =  \Big [ 0.0392;\; 0.0144; \; 0.0183; \;0.0058;  \;1.4665 \Big ] \times 10^{-6} . \label{eq:A0}
\end{eqnarray}
\end{subequations}
Superscripts were omitted for the globally identifiable parameters and initial conditions, since they take the same value in both sets.
As before, we integrate the model with the above ``exact" parameters and initial conditions, 
and then add a prescribed level of noise to the numerical solution $D(t)$. This polluted solution is then taken as 
the reported data and used to obtain the range
of variation of the model predictions. 
 
\begin{figure}[htbp]
 \includegraphics[scale=0.125]{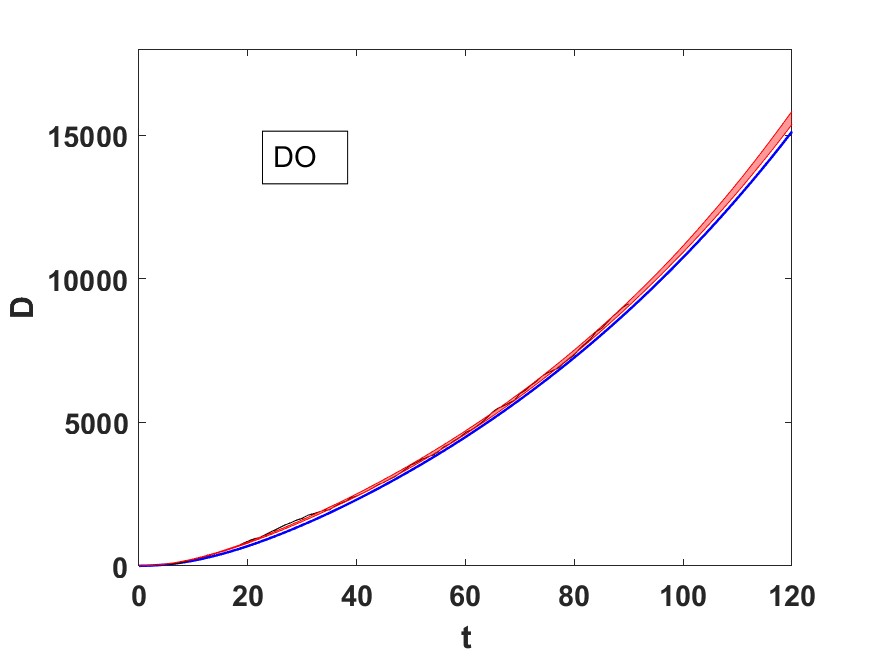} 
 \includegraphics[scale=0.125]{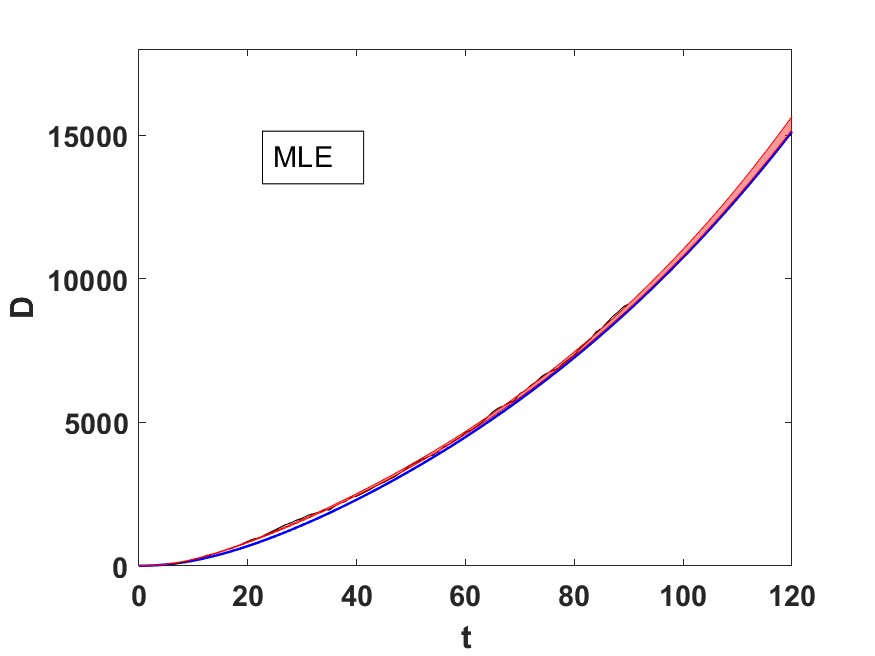} 
 \includegraphics[scale=0.125]{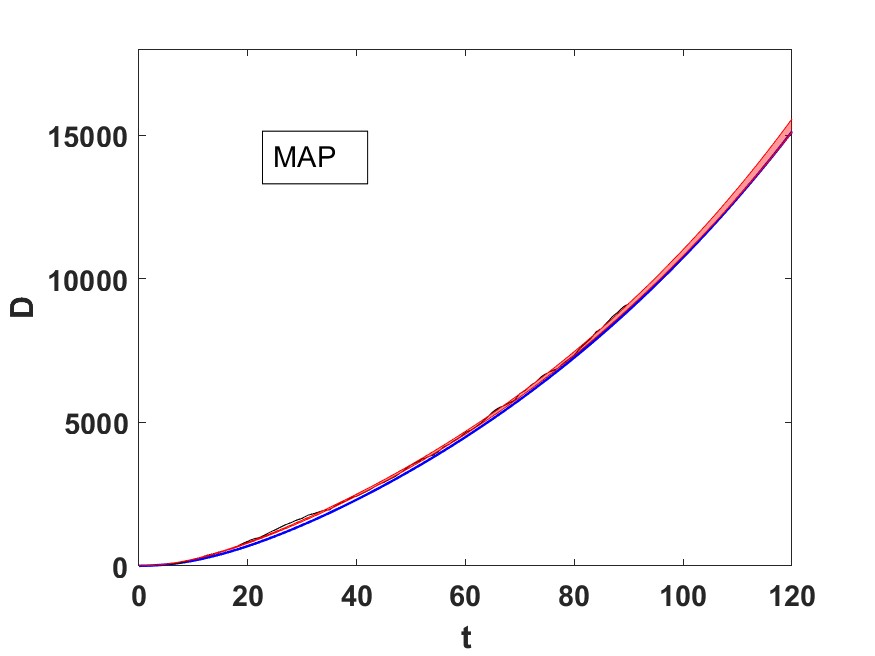}
\includegraphics[scale=0.125]{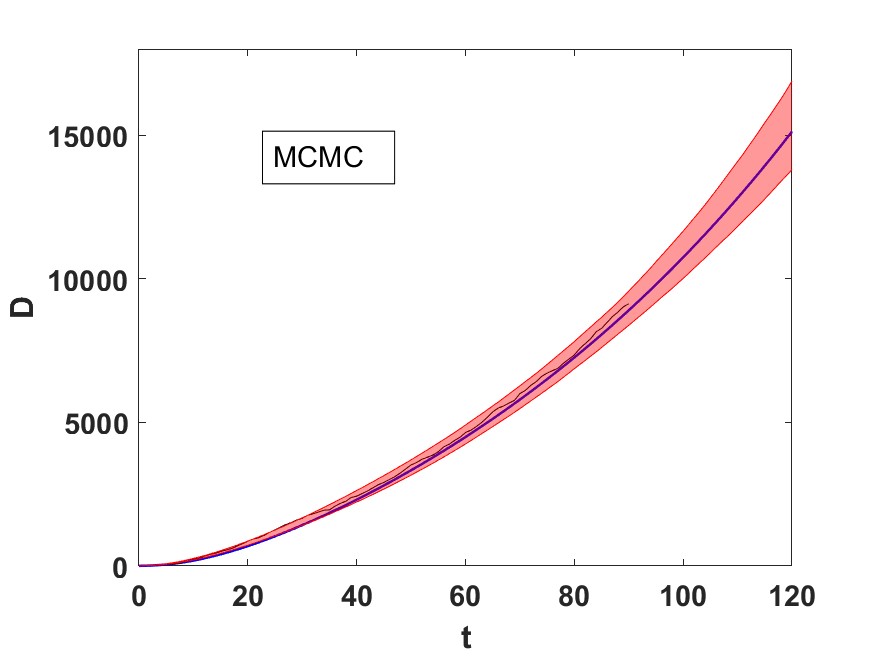}
 \includegraphics[scale=0.125]{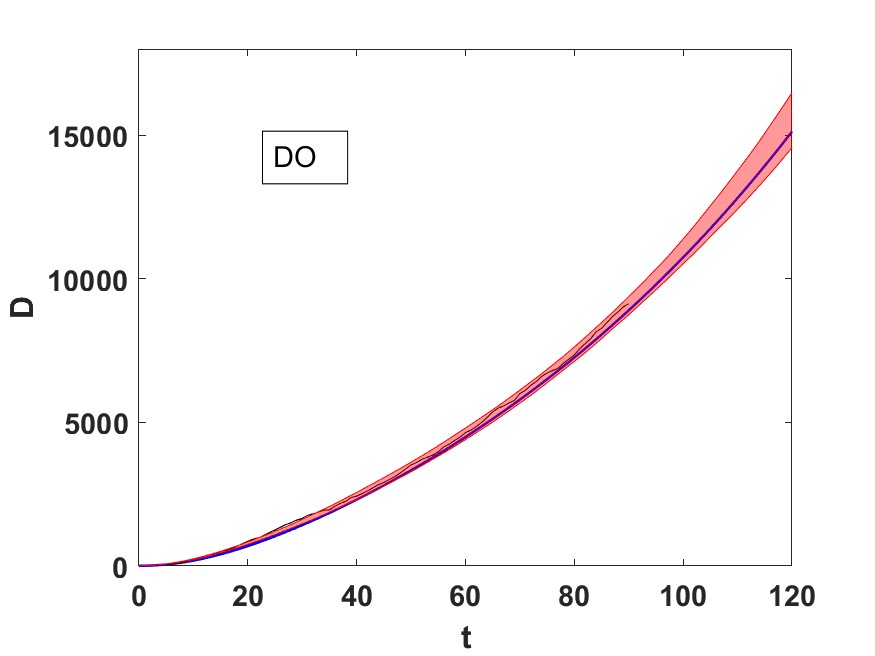}  
 \includegraphics[scale=0.125]{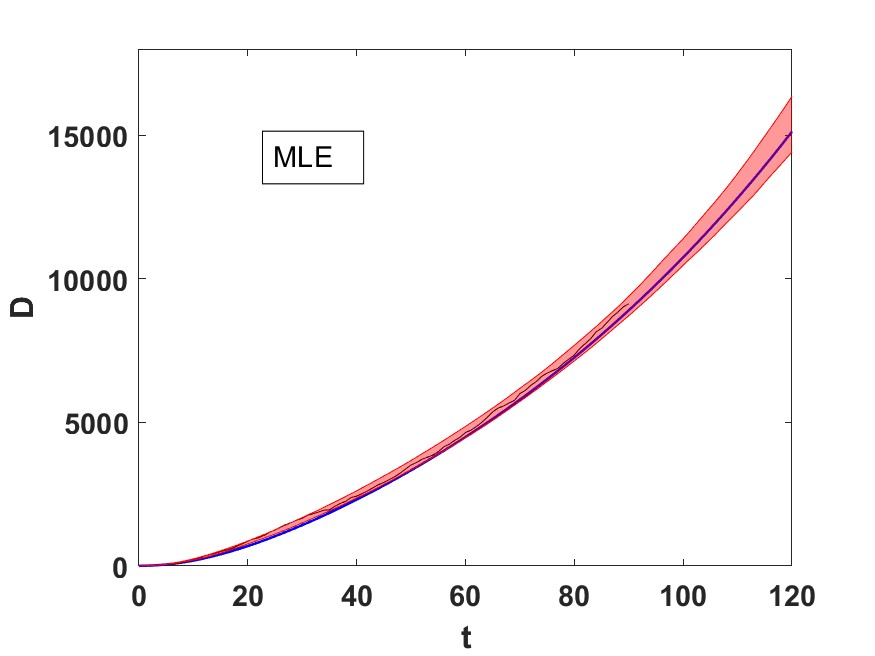} 
 \includegraphics[scale=0.125]{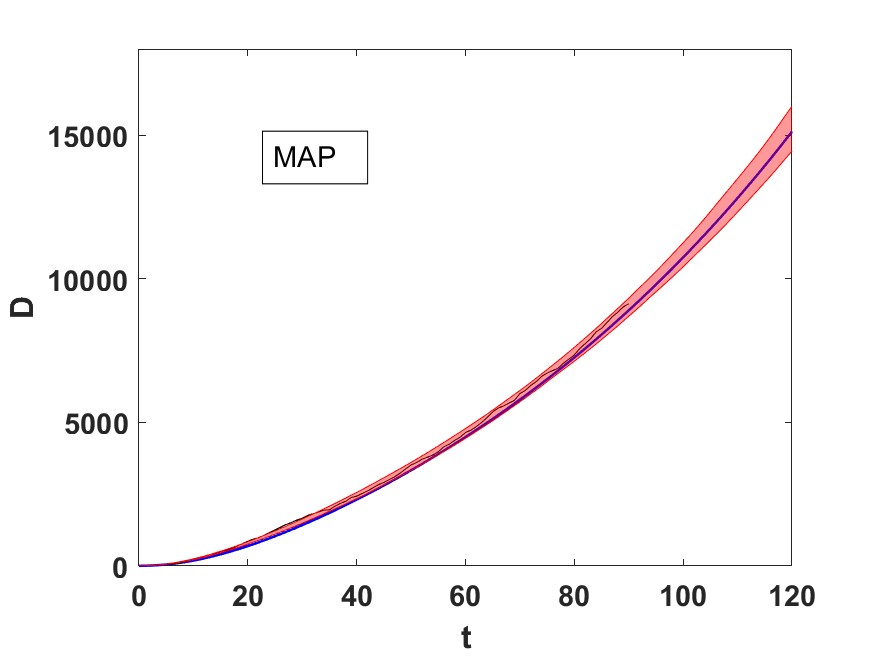} 
\includegraphics[scale=0.125]{CSF_pred_4thA_Normal_NP_90_small_DeltaNoise30_Delta_M6_D}
\caption{EAIHRD model.  Predicted number of daily fatalities $D(t)$. Small bounds for $\tilde{A}(0)$.
Synthetic data noise level $\hat{\sigma}_{\varepsilon} = 0.30$.
Top row: DO, MLE, MAP predictions with bootstrapping noise $\sigma_{\varepsilon} = 0.05$, and MCMC prediction.
Bottom row: DO, MLE, MAP predictions with bootstrapping noise $\sigma_{\varepsilon} = 0.20$, and MCMC prediction.
 Initial conditions for $D$ and its derivatives are treated as unknowns. 
 Algorithm training period $\tTrain=90$.}
 \label{f:CSF_pred_4thA_D_small}\end{figure}

\begin{figure}[htbp]
 \includegraphics[scale=0.125]{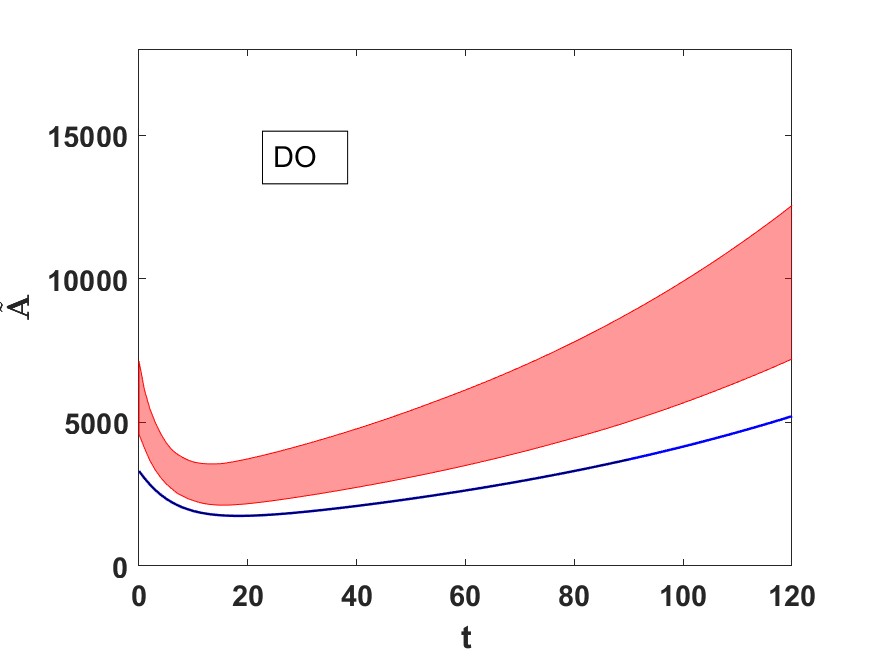} 
 \includegraphics[scale=0.125]{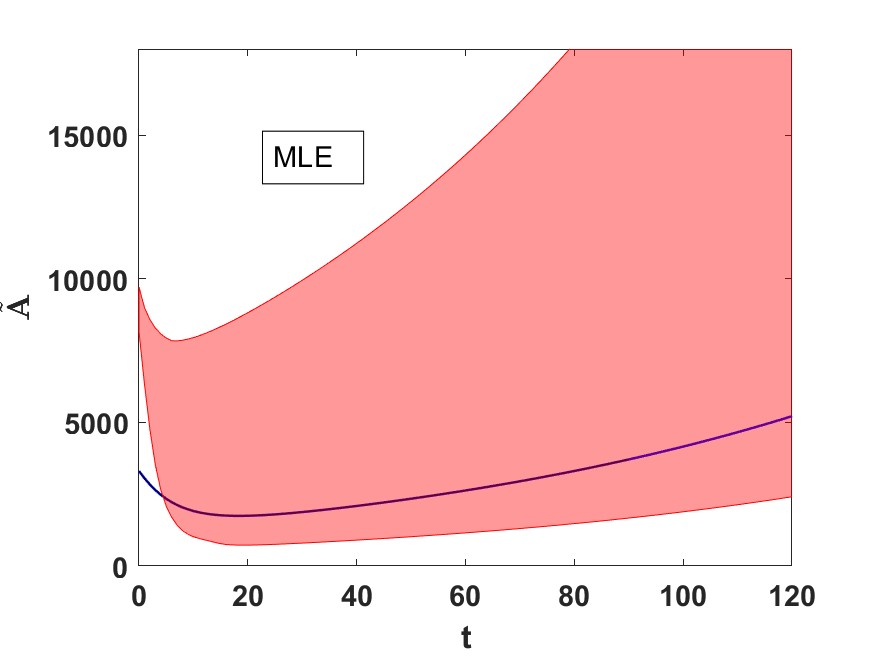} 
  \includegraphics[scale=0.125]{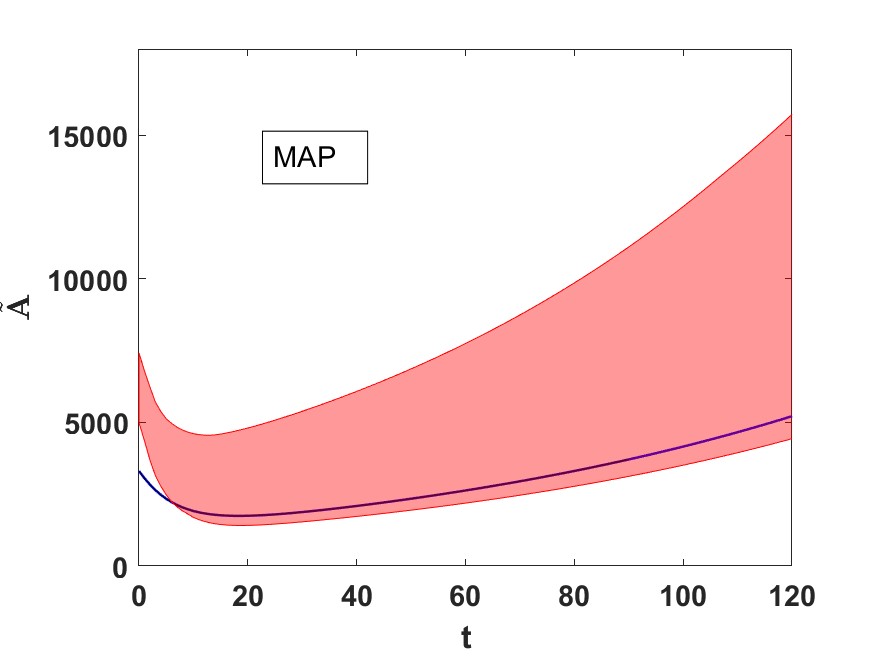} 
\includegraphics[scale=0.125]{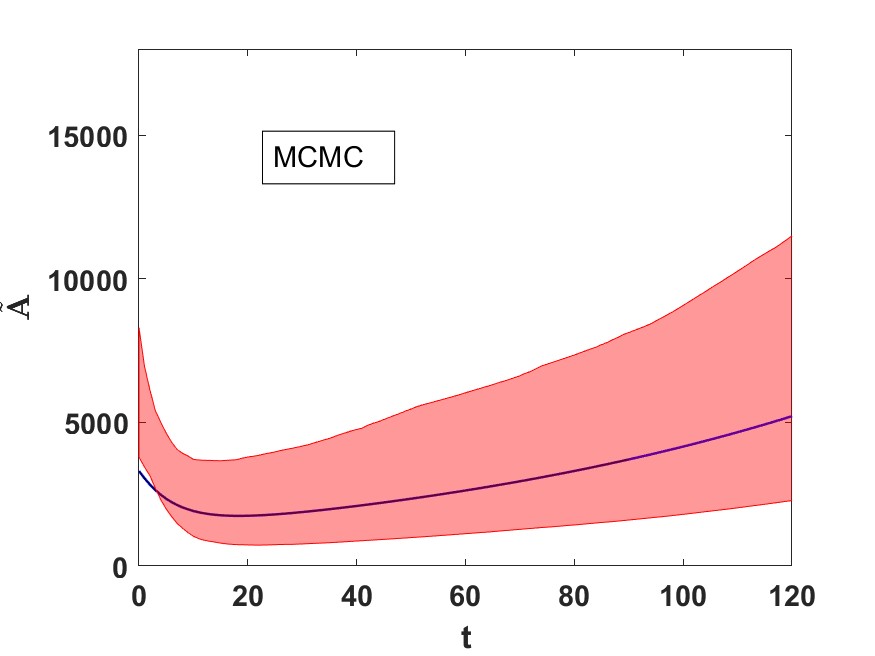}
  \includegraphics[scale=0.125]{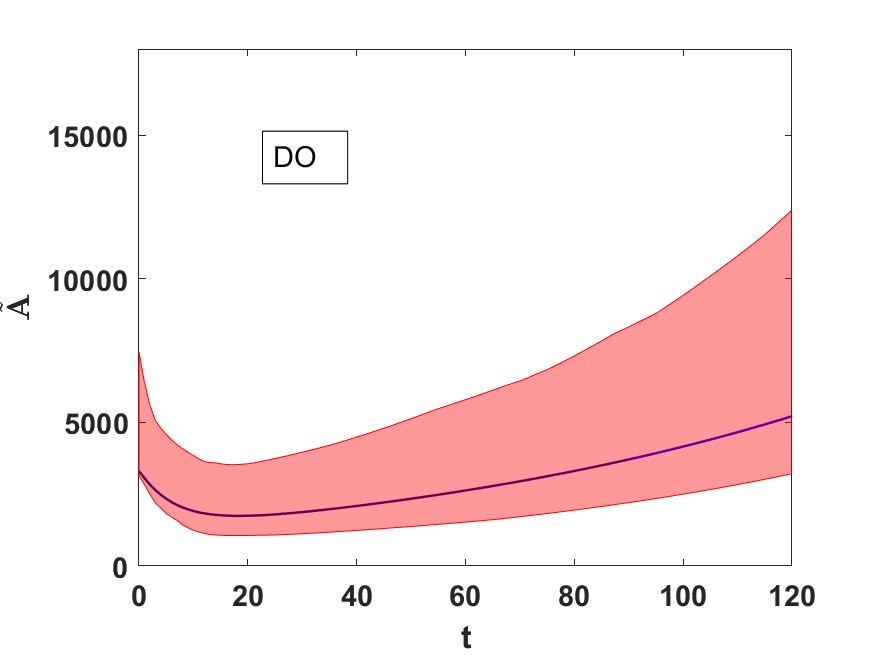} 
   \includegraphics[scale=0.125]{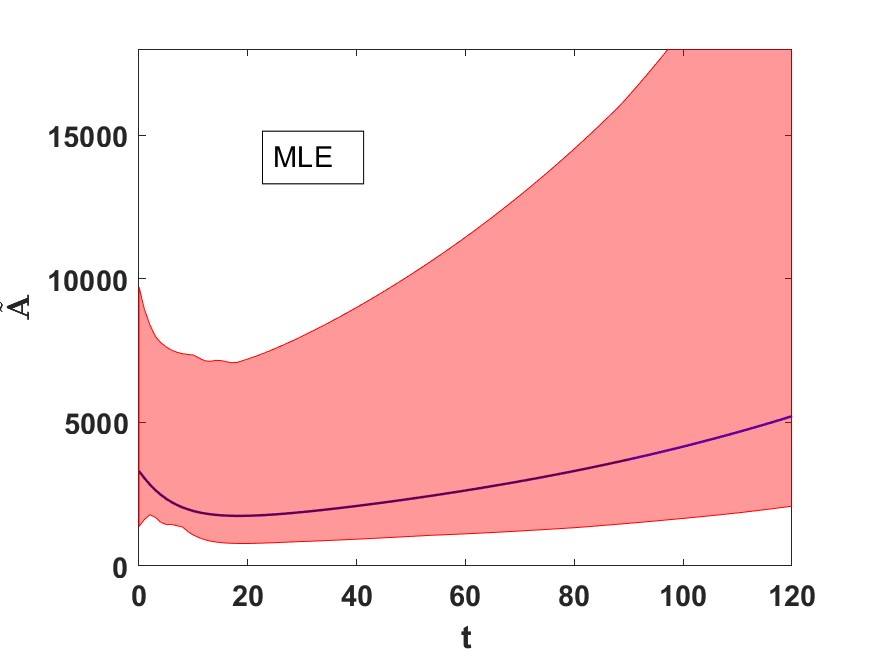} 
 \includegraphics[scale=0.125]{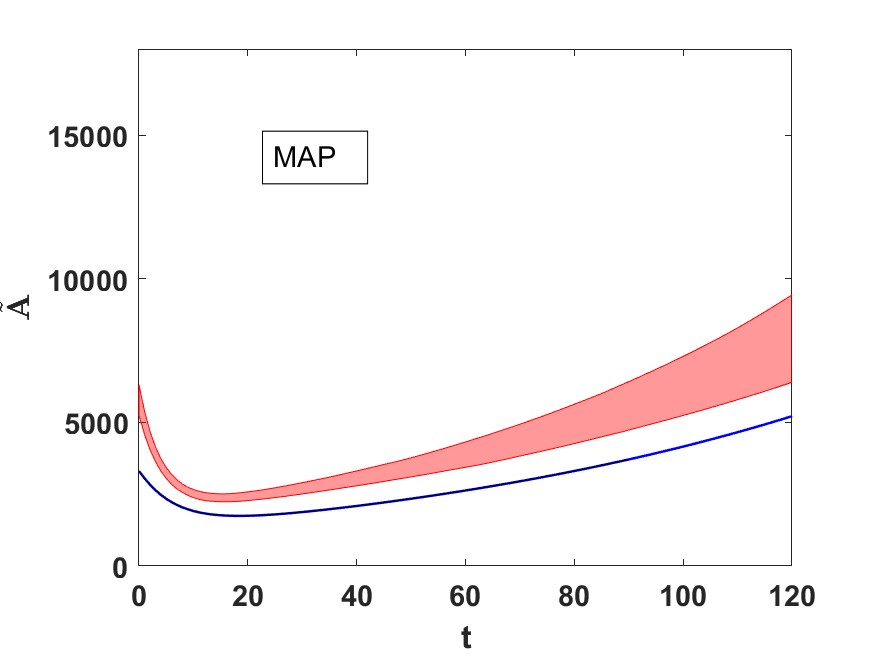} 
\includegraphics[scale=0.125]{CSF_pred_4thA_Normal_NP_90_small_DeltaNoise30_Delta_M6_A}
\caption{EAIHRD model. 
 Predicted number of asymptomatic individuals $\tilde{A}(t)$. Same setup as in Fig.~\ref{f:CSF_pred_4thA_D_small}.
}
\label{f:CSF_pred_4thA_A_small}\end{figure}

In the first test, we treat all parameters and initial conditions as unknowns to be estimated. 
From the analysis in Sec.~\ref{sec:IdentifyFokas}, several parameters and $\tilde{A}(0)$ are locally identifiable with two possible values. 
We enforce a small bounding interval on $\tilde{A}(0)$ to ensure that only the value 
$\tilde{A}^{(1)}(0)$,  cf. Eq.~(\ref{eq:A0}), is possible if the reported data are noise-free. 
That is, the system is globally identifiable if the
reported data do not have noise. 
The bounds we used were:
\begin{eqnarray*}
 F \in [0.05, 0.7  ]; \quad r_1 \in [ 0.05, 0.4 ];    \quad R_2 \in  [0.05, 0.7  ]; \quad R_3 \in  [0.04, 0.3  ]; \\
C_1 \in [1, 60];  \quad C_2 \in  [100, 250  ]; \quad \alpha \in  [2, 7] \times  10^{-4}; \\
 D(0) \in [0.1, 0.8  ] \times 10^{-7}; \quad D'(0) \in  [0.05, 0.3  ] \times 10^{-7}; \quad 
D''(0) \in  [0.1, 0.4  ] \times 10^{-7}; \\  D'''(0) \in  [0.02, 0.08  ] \times 10^{-7}; \quad 
 \tilde{A}(0) \in  [0.2, 5] \times 10^{-7}
\end{eqnarray*}
We also treat $D(0)$ as an unknown,  corresponding to a scenario that we do not trust fully the reported data. 
In our numerical results, there appears no noticeable difference whether we fix $D(0)$ or not,
which is mainly due to the fact that the estimate for $D(0)$ is very accurate.
We estimated the parameters with a data noise level $\hat{\sigma}_{\varepsilon}=0.30$.
The predicted fatalities time series $D(t)$
is shown in Fig.~\ref{f:CSF_pred_4thA_D_small}.
The panels in the top row include MCMC result and  the results for DO, MLE, and MAP with bootstrapping noise 
$\sigma_{\varepsilon}=0.05$. Those in the bottom 
row are for  $\sigma_{\varepsilon}=0.20$.
All four estimators give reasonable predictions for $D(t)$.
Figure~\ref{f:CSF_pred_4thA_A_small}, on the other hand
shows  model predictions for a non-observable, the number of asymptomatics
$\tilde{A}(t)$. The panels are as in Fig. ~\ref{f:CSF_pred_4thA_D_small}.
The results shown do not answer unequivocally
whether the prediction of $\tilde{A}(t)$ improves with increasing artificial noise in
bootstrapping.

\begin{figure}[htbp]
 \includegraphics[scale=0.125]{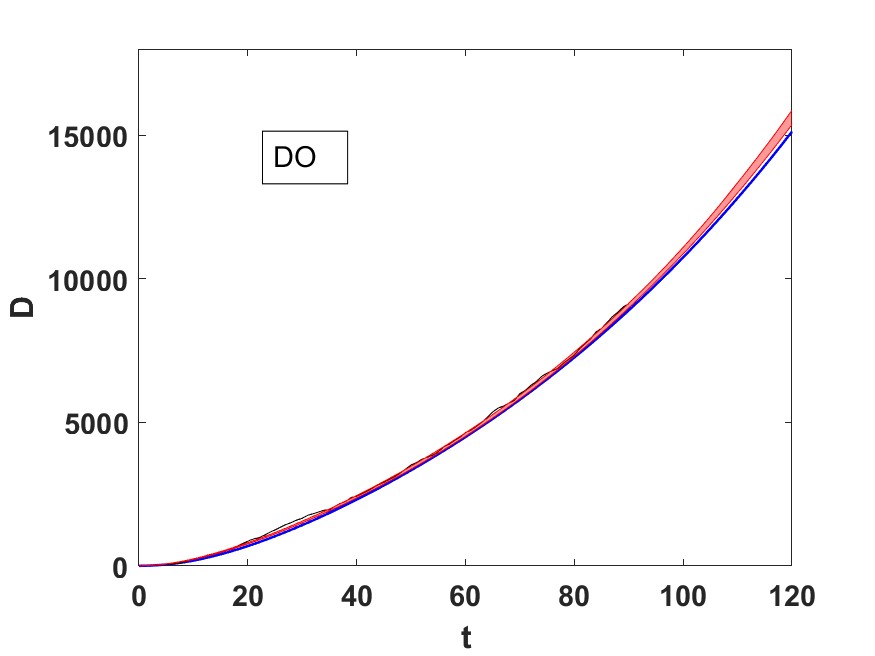} 
 \includegraphics[scale=0.125]{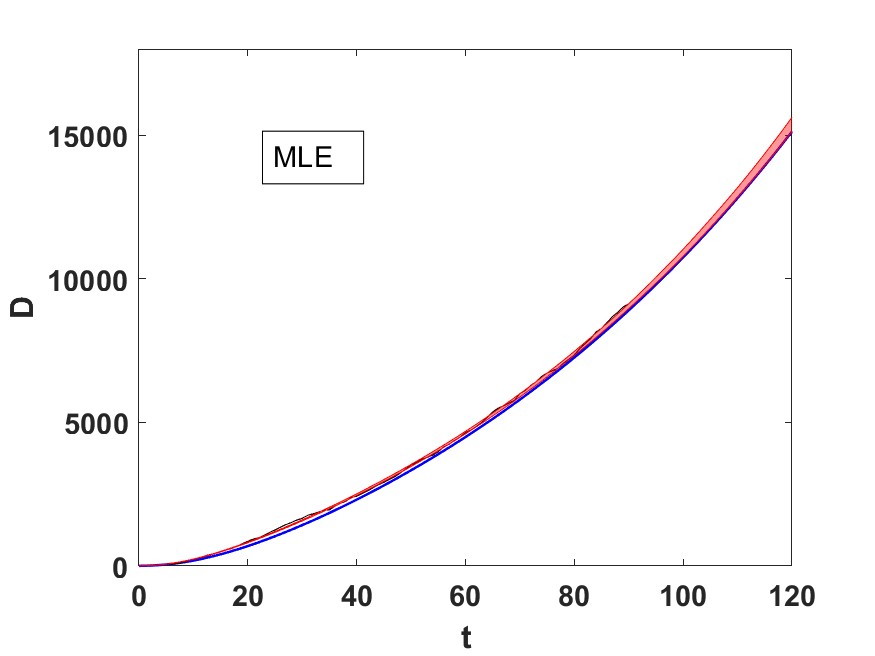}
  \includegraphics[scale=0.125]{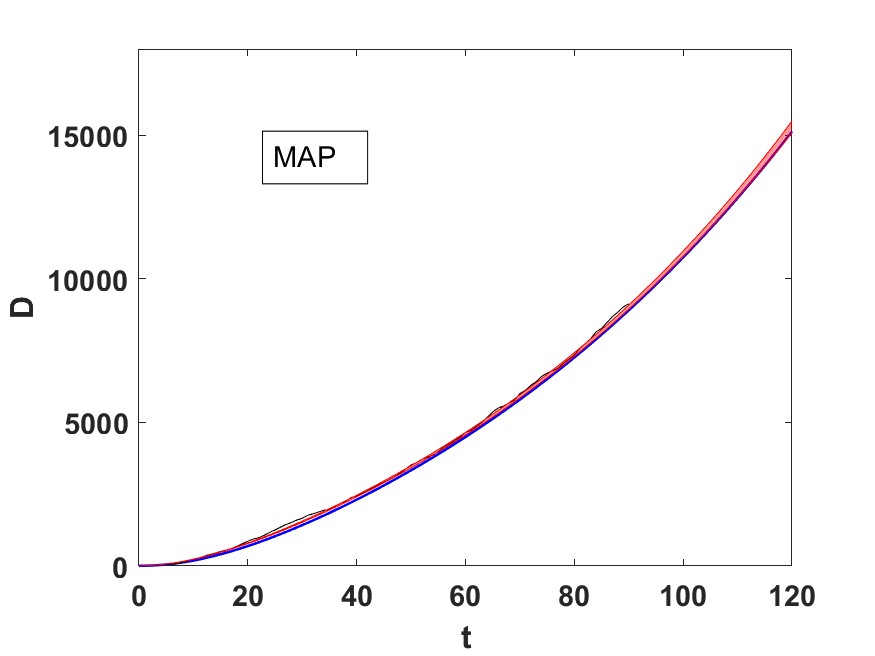}
\includegraphics[scale=0.125]{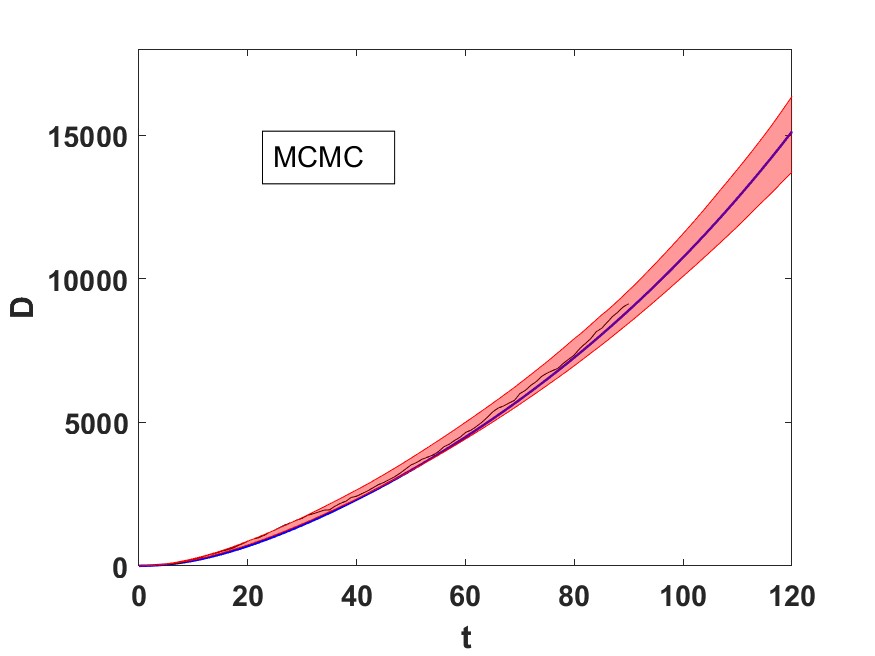}
  \includegraphics[scale=0.125]{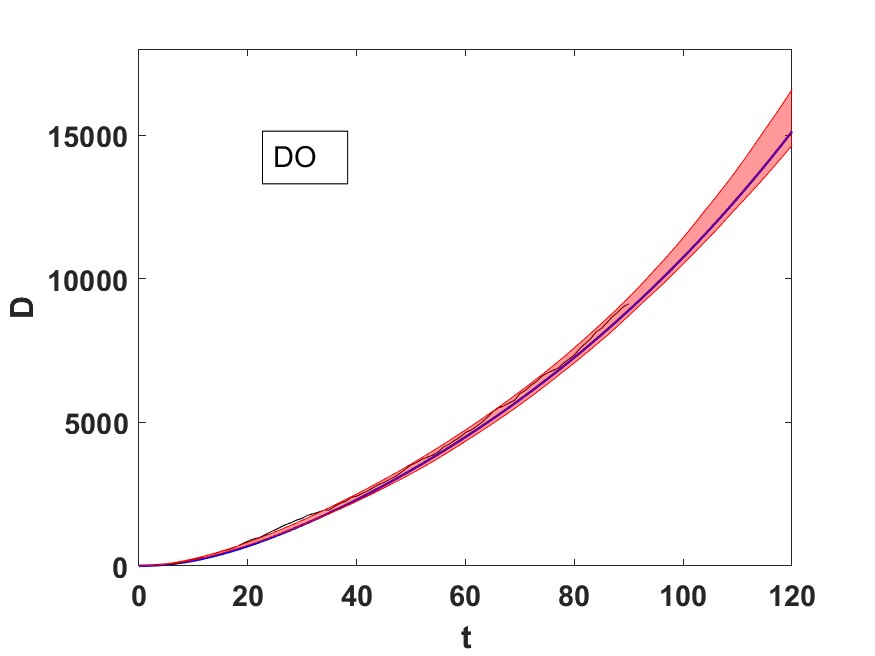}
 \includegraphics[scale=0.125]{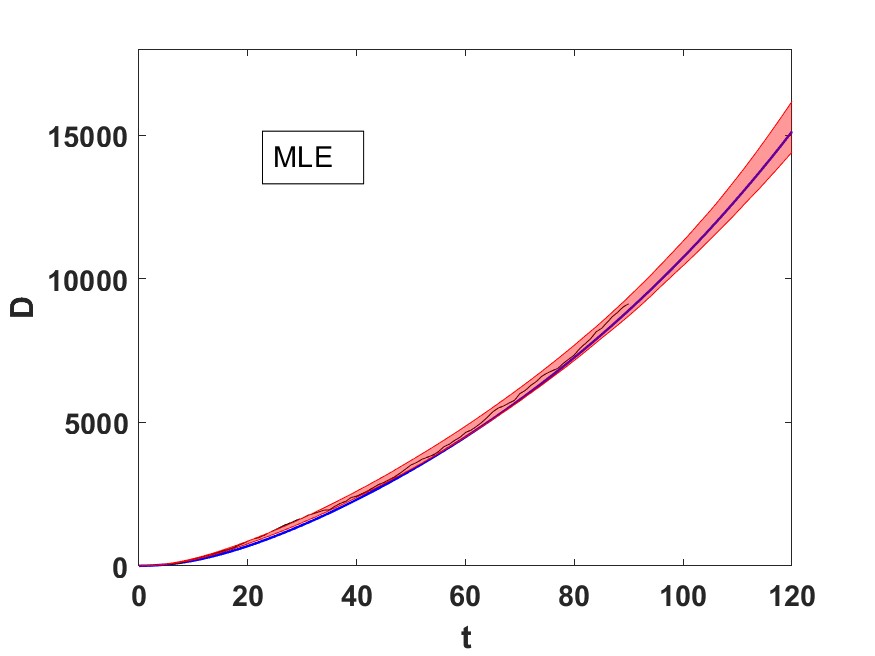} 
 \includegraphics[scale=0.125]{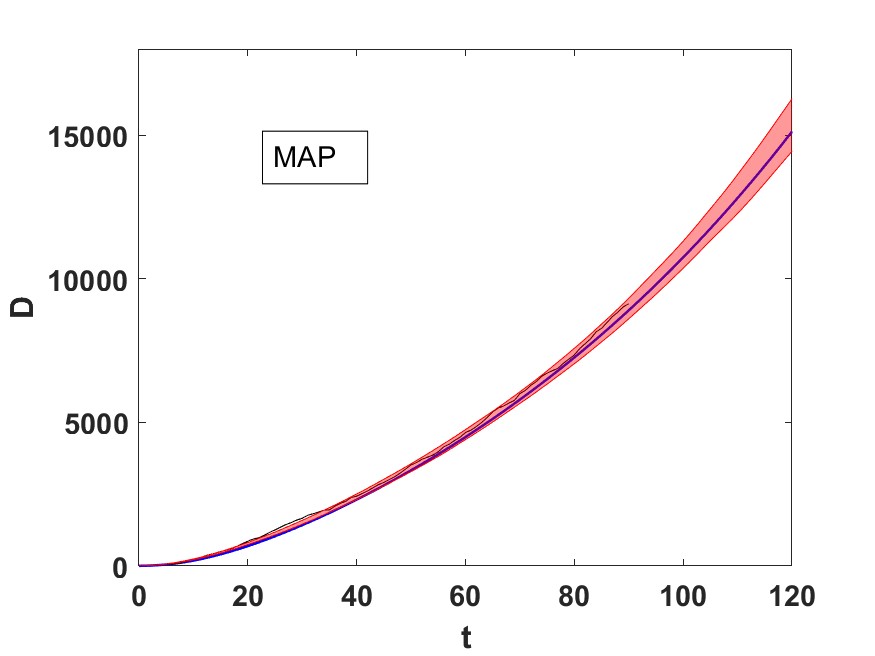} 
\includegraphics[scale=0.125]{CSF_pred_4thA_Normal_NP_90_DeltaNoise30_Delta_M6_D}
\caption{EAIHRD model.  Predicted number of daily fatalities $D(t)$. Initial condition $\tilde{A}(0)$ given.
Synthetic data noise level $\hat{\sigma}_{\varepsilon} = 0.30$
Top row: DO, MLE, MAP predictions with bootstrapping noise $\sigma_{\varepsilon} = 0.05$, and MCMC predictions.
Bottom row: DO, MLE, MAP predictions with bootstrapping noise $\sigma_{\varepsilon} = 0.20$, and MCMC predictions.
 Initial conditions for $D$ and its derivatives are treated as unknowns. 
 Algorithm training period $\tTrain=90$.
 } \label{f:CSF_pred_4thA_D_fixA0}\end{figure}

\begin{figure}[htbp]
 \includegraphics[scale=0.125]{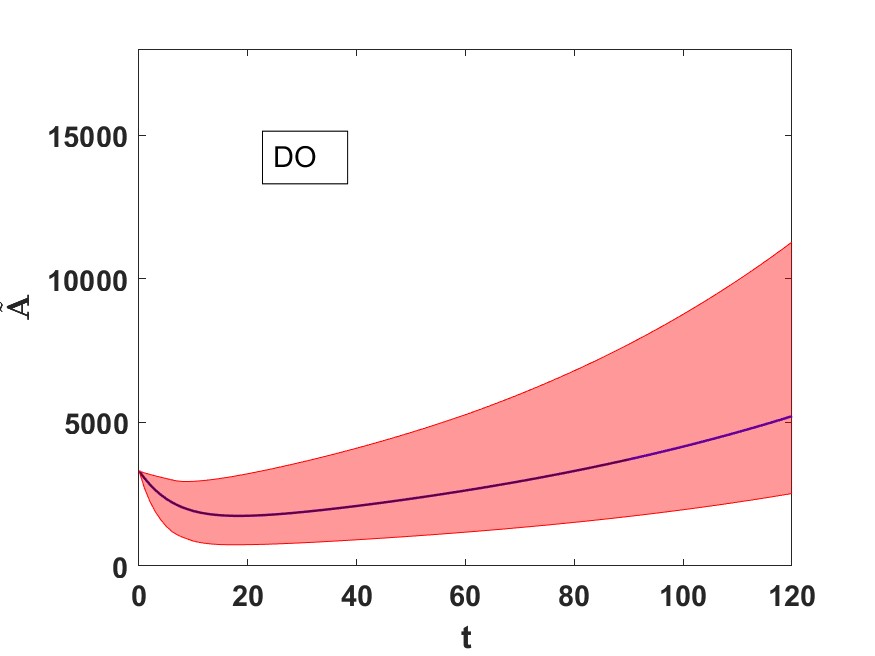} 
 \includegraphics[scale=0.125]{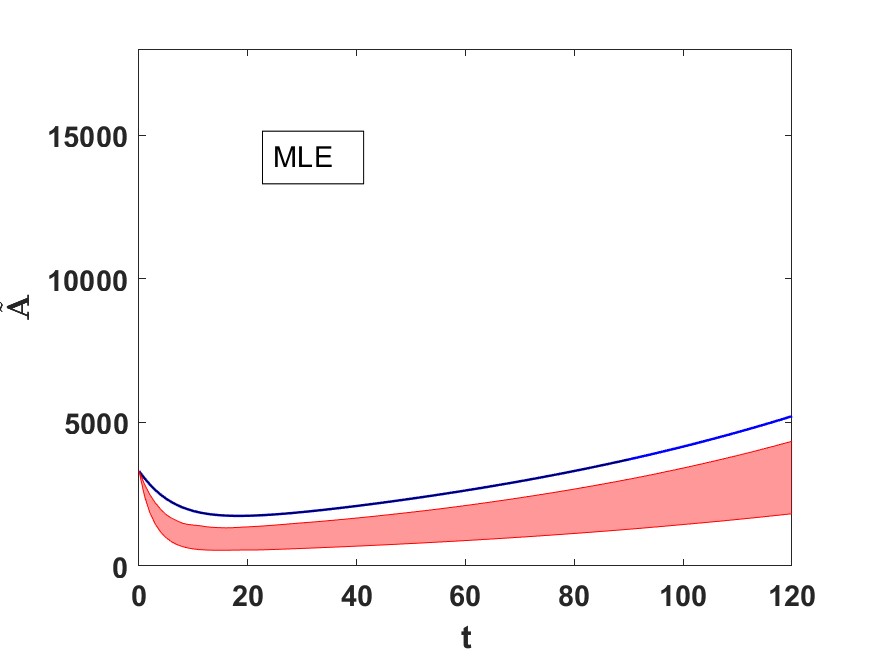} 
  \includegraphics[scale=0.125]{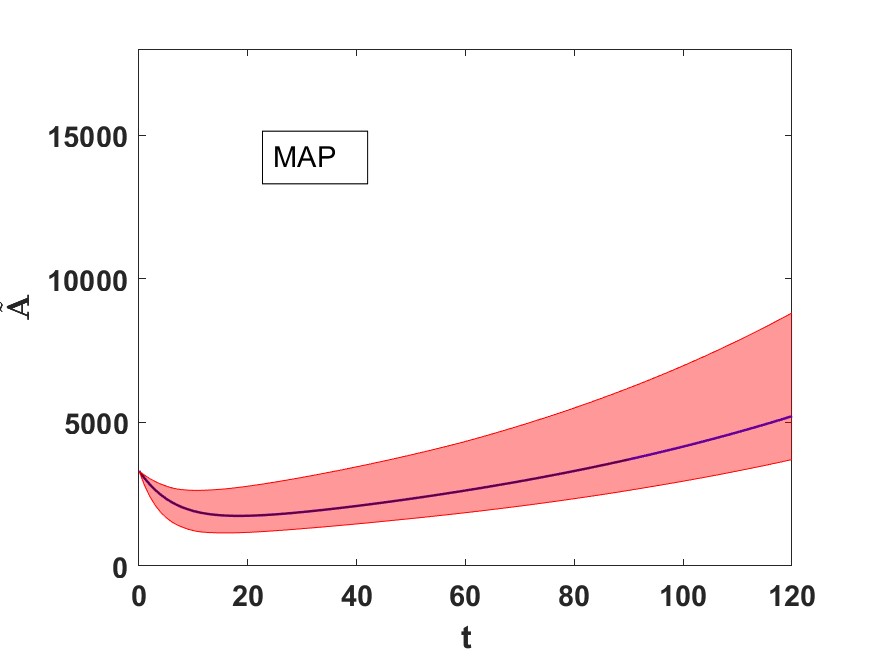} 
\includegraphics[scale=0.125]{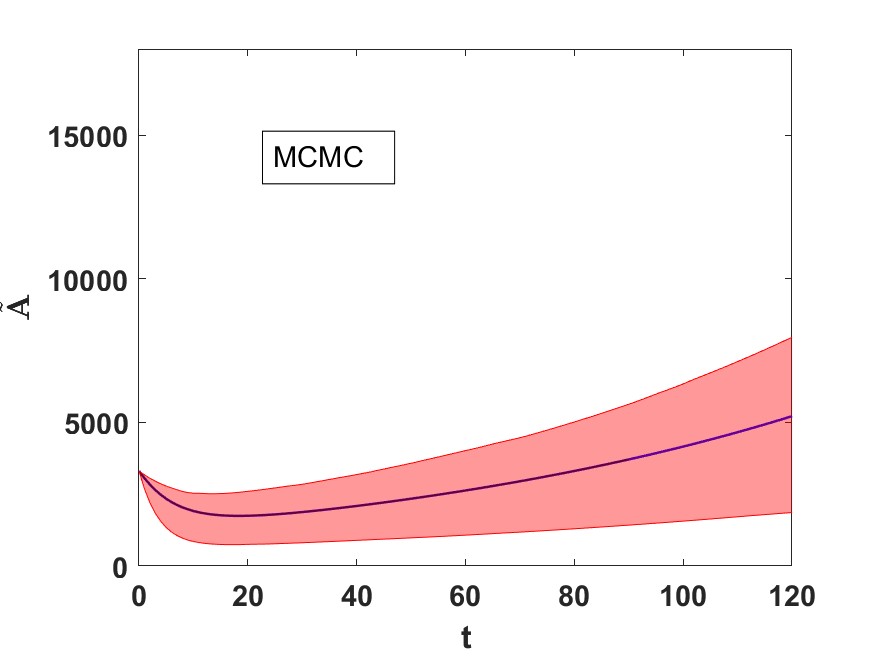}
  \includegraphics[scale=0.125]{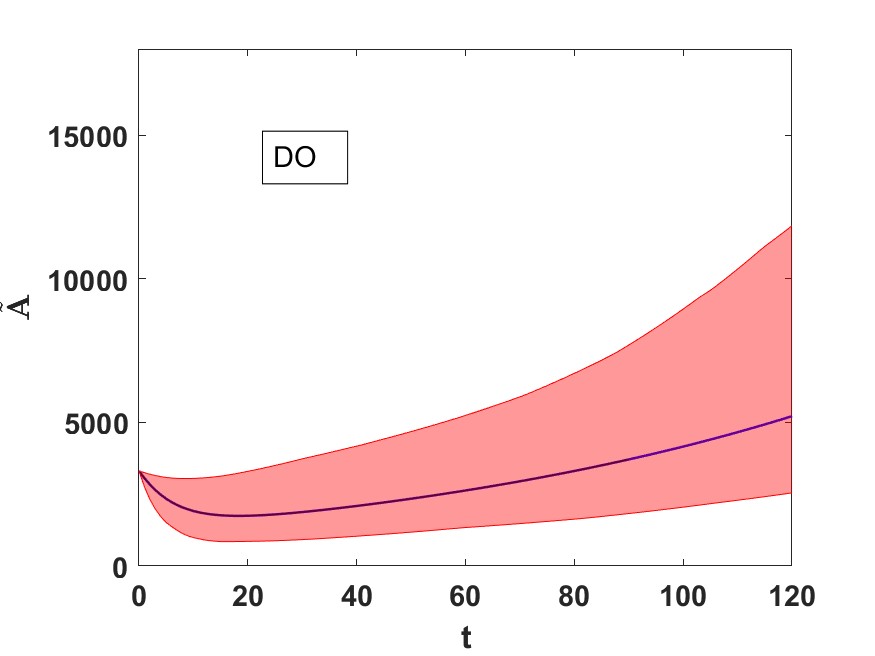}
 \includegraphics[scale=0.125]{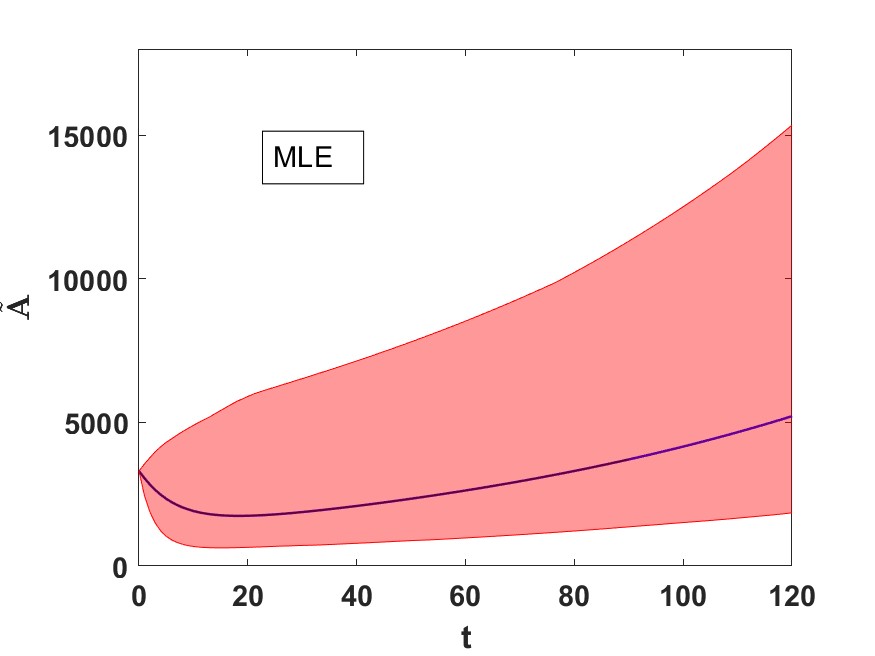} \
 \includegraphics[scale=0.125]{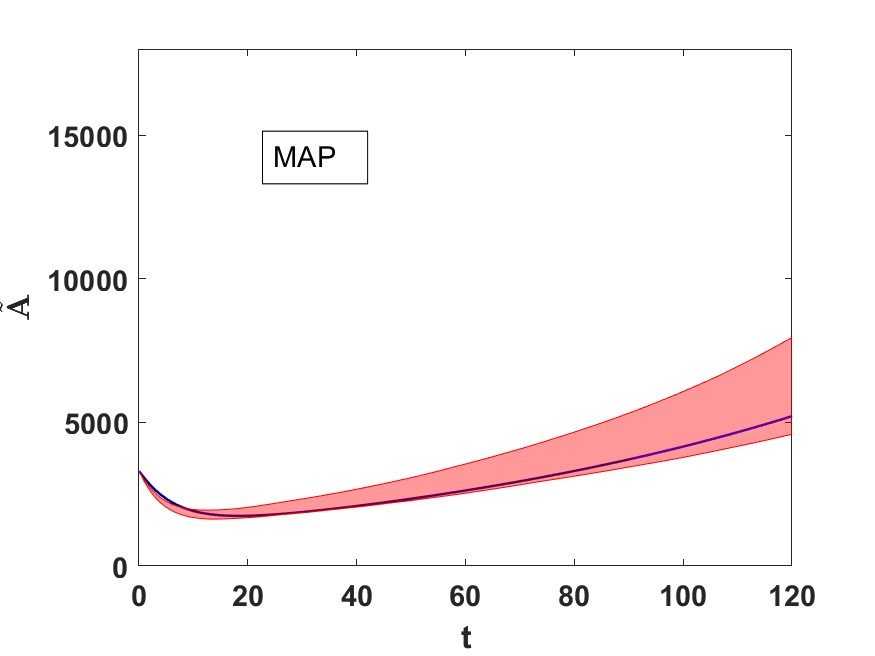} 
\includegraphics[scale=0.125]{CSF_pred_4thA_Normal_NP_90_DeltaNoise30_Delta_M6_A}
\caption{EAIHRD model.  Predicted number of asymptomatic individuals $\tilde{A}(t)$. 
Same setup as in Fig.~\ref{f:CSF_pred_4thA_D_fixA0}
 } \label{f:CSF_pred_4thA_A_fixA0}\end{figure}

For the second test case, we assume that $\tilde{A}(0)$ is given, so the system becomes globally identifiable.  The resulting model predictions for the number of fatalities $D(t)$ are shown in Fig.~\ref{f:CSF_pred_4thA_D_fixA0}. Compared to Fig.~\ref{f:CSF_pred_4thA_D_small}, where we used small bounds for $\tilde{A}(0)$, there is no noticeable difference in assuming $\tilde{A}(0)$ is known.
Similarly, the results for the number of asymptomatics $\tilde{A}(t)$ are shown in Fig.~\ref{f:CSF_pred_4thA_A_fixA0} (with known $\tilde{A}(0)$) and Fig.~\ref{f:CSF_pred_4thA_A_small} (with a small bound for $\tilde{A}(0)$). Now, optimizers DO, MAP perform better when the noise level is $\sigma_{\varepsilon} =0.05$, MLE and MAP perform better when the noise level is  $\sigma_{\varepsilon} =0.20$, and MCMC performs better when $\tilde{A}(0)$ is known.

\begin{figure}[htbp]
 \includegraphics[scale=0.25]{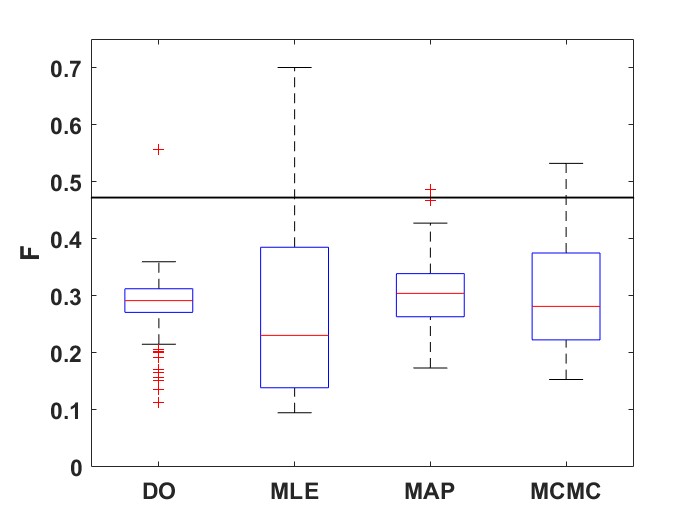} 
 \includegraphics[scale=0.25]{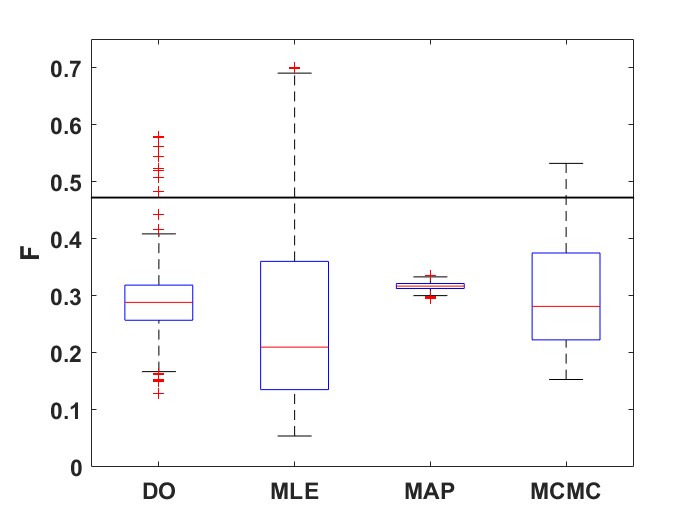} 
 \includegraphics[scale=0.25]{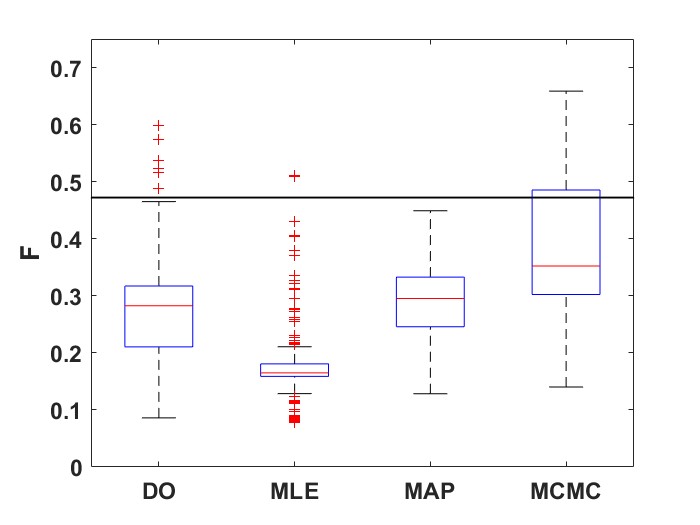}  
 \includegraphics[scale=0.25]{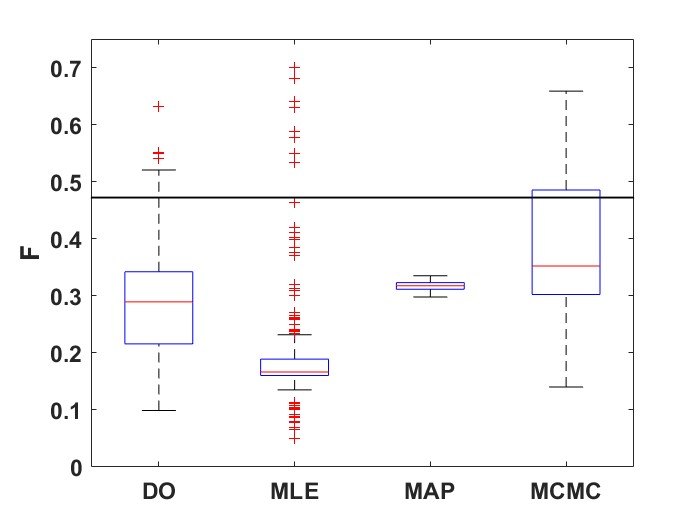} 
\caption{EAIHRD model. Box plot for parameter $F$. Top:  $\tilde{A}(0)$ is treated as a parameter, but small bounds were enforced during
the estimation procedure. Bottom: $\tilde{A}(0)$ is given. Left: noise level $\sigma_{\varepsilon} = 0.05$ in bootstrapping;
Right: noise level $\sigma_{\varepsilon} = 0.20$ in bootstrapping.
$\hat{\sigma}_{\varepsilon} = 0.30$ noise in data and  $\tTrain=90$.
 } \label{f:CSF_F_boxplot}\end{figure}

As an example of the accuracy of the estimated parameters,
and as a means to interpret the result shown in the time-series figures,
we show the box plotted estimated $F$ 
in Fig.~\ref{f:CSF_F_boxplot}.
We introduced noise in the data with $\hat{\sigma}_{\varepsilon}=0.30$, and 
noise in bootstrapping with $\sigma_{\varepsilon}=0.05, 0.20$.
The top-row panels were obtained with $\tilde{A}(0)$ bounded in a small interval,
whereas the bottom row with $\tilde{A}(0)$ fixed.
In general as the bootstrapping noise level increases,
so does the interquartile range. But for this particular model, the interquartile range actually decreases for the MAP predictions (cf Figs.~\ref{f:CSF_pred_4thA_A_small} and ~\ref{f:CSF_pred_4thA_A_fixA0}).  Irrespective of the
bootstrapping noise,  the exact $F$ does not lie within the $25-75\%$ interval
predicted by the three point estimators, 
with the Bayesian MCMC estimator clearly outperforming them,
although solely in the case where the $\tilde{A}(0)$ is given.

As shown in Fig.~\ref{f:CSF_F_boxplot}, the $F$ estimates are not very  good.
Even when $\tilde{A}(0)$ is specified, leading to a globally identifiable system, 
only the MCMC estimate seems acceptable.
They are certainly much worse than the $\sigma$ estimates obtained in the SEIR model,  cf. Fig.~\ref{f:FixS0_boxplots_sigma}.
This arises because the observable $D(t)$ is much less sensitive to $F$ (than $I(t)$ on $\sigma$),
as indicated by the gap in the eigenvalues of the Hessian.
When $\tilde{A}(0)$ is given, the eigenvalues of Hessian are: 

\begin{center}
\begin{tabular}{*{6}{c}}
$2.82 \times 10^{21}$ & $2.79 \times 10^{17}$ & $8.38 \times 10^{14}$ & $4.55 \times 10^{13}$ &
$1.82 \times 10^6$ \\  $2.43 \times 10^2$ & $30.11$ & $0.25$ & $0.11$ & $1.20\times 10^{-3}$ & $1.34 \times 10^{-5}$
\end{tabular}
\end{center}

Hence, the eigengap is much larger than the eigengap of the Hessian
in the SEIR mode,  implying that $D(t)$ is much less sensitive to parameter $F$,
which leads to a worse estimate for $F$  in EAIHRD than $\sigma$ in SEIR 
(Figs~\ref{f:FixS0_boxplots_sigma} vs~\ref{f:CSF_F_boxplot}). 
%

\section{Conclusions}
\label{sec:Concl}

When faced with a process such as the spread of an epidemic, mathematical modelers have to make several decisions, some of which are more straightforward than others. For instance, from the early stages of the COVID-19 pandemic, modelers became aware that asymptomatic and presymptomatic infections played a role \cite{Mexico}. Hence, classes of asymptomatic, presymptomatic people, or both were included in epidemiological models. On the other hand, complications may arise due to the limited length and quality of available time series data. Specifically, having data for only the initial stages of an epidemic is not enough to accurately estimate epidemic parameters \cite{chowell2017} (see also~\cite{sauer2021}). Also, there were questions regarding the reliability of the initial data for COVID-19 \cite{backcasting}. Therefore, even when data do exist, whether they can be trusted and deemed sufficient for parameter estimation or not is not always clear. Other modeler choices relate to the fitting methods that are used, namely whether a frequentist (e.g., least-squares method) or Bayesian approach is preferable.

In this study, we compare several parameter estimation methods -- deterministic optimization, maximum likelihood, maximum a posteriori estimation, MCMC -- using synthetic data. This is  a  conscious choice enabling us to focus on known sources of noise and to fully control data quantity and quality. We also consider two different models, the well-studied, simpler SEIR (susceptible, exposed, infectious, recovered) model and a more involved model, which also accounts for
(the important especially during the COVID-19 pandemic) asymptomatic, $A(t)$, hospitalized, $H(t)$, and deceased, $D(t)$ populations. Adding to an already substantial literature of parameter identifiability studies \cite{chowell2017, tuncer2018, zhang2021}, our work focuses on practical identifiability. We  investigate the effect of noise level in the data and in bootstrapping, as well as the role of initial conditions in the estimated parameters and how these
affect local vs. global identifiability both theoretically (structurally)
and practically. 

In some cases, we  obtain clear answers, such as that the MCMC estimator, although time-consuming to implement, is superior to the others we considered. 
While this may not be a ``universal outcome'', it was systematic enough in 
our findings that it is worth reporting. 
In other cases, the outcomes of our numerical experiments, such as whether fitting or fixing the initial conditions improves practical identifiability, are more nuanced. Specifically, in the SEIR model, when the initial condition of $S(t)$, which is not part of the data, was fixed, it improved parameter estimates. In the more involved model, fixing the initial condition for $\tilde{A}(t)$, which is not directly measured either, (or/and the initial condition for fatalities) 
did not significantly improve parameter estimates.  
We also corroborated earlier findings about the quality of
parameter estimates when training changes from pre-peak to post-peak 
in time-series scenarios. Additionally, we illustrated the role
that noise can play in ``annulling'' global identifiability results
in cases where there is lack of sensitivity of an observable on
some parameters or initial conditions. We did accordingly accentuate
the importance of appreciating that sensitivity via the eigenvalue spectral gap.

We believe that studies of this kind shed light into key issues in parameter estimation for applications that have real-world importance and are worthwhile to expand
upon both in the case of simpler, as well as in that of more complex (and also
spatially dependent~\cite{byrne24}) models. 



\section*{Acknowledgements}

Support from C3.ai Inc. and the Microsoft Corporation, as well as the NSF (grant DMS-1815764 to Z.R.) is acknowledged. J.C.-M. acknowledges support from the EU (FEDER program 2014–2020) through MCIN/AEI/10.13039/501100011033 (under the projects PID2020-112620GB-I00 and PID2022-143120OB-I00).

\clearpage \newpage

\end{document}